\documentclass[journal,letterpaper]{IEEEtran}
\usepackage{blindtext,cite,graphicx,booktabs,multirow,mathtools,amssymb,xcolor}
\usepackage[T1]{fontenc}
\usepackage[hidelinks]{hyperref}
\hypersetup{
    colorlinks=true,       
    linkcolor=violet,          
    citecolor=green,        
    urlcolor=blue           
}
\usepackage[font=small]{caption}
\usepackage[subrefformat=parens,labelformat=parens,skip=0pt]{subcaption}

\usepackage[normalem]{ulem} 
\newcommand{\overbar}[1]{\mkern 1.5mu\overline{\mkern-1.5mu#1\mkern-1.5mu}\mkern 1.5mu}

\newcommand{\ac}{\ifthenelse{\boolean{eVersion} \or \boolean{instructorVersion}}{red}{dashed}}
\newcommand{\rc}{\ifthenelse{\boolean{eVersion} \or \boolean{instructorVersion}}{black}{solid}}
\newcommand{\gc}{\ifthenelse{\boolean{eVersion} \or \boolean{instructorVersion}}{gray}{solid gray}}
\newcommand{\acrg}{\ifthenelse{\boolean{eVersion} \or \boolean{instructorVersion}}{red}{gray}}
\newcommand{\acrdg}{\ifthenelse{\boolean{eVersion} \or \boolean{instructorVersion}}{red}{dark gray}}

\newcommand{\C}{\ensuremath{\mathbb{C}}}
\newcommand{\Z}{\ensuremath{\mathbb{Z}}}

\DeclareFontFamily{U}{mathx}{\hyphenchar\font45}
\DeclareFontShape{U}{mathx}{m}{n}{
      <5> <6> <7> <8> <9> <10>
      <10.95> <12> <14.4> <17.28> <20.74> <24.88>
      mathx10
      }{}
\DeclareSymbolFont{mathx}{U}{mathx}{m}{n}
\DeclareFontSubstitution{U}{mathx}{m}{n}
\DeclareMathAccent{\widecheck}{0}{mathx}{"71}

\newcommand{\Frac}[2]{{{#1}/{#2}}}  

\newcommand{\ds}{\displaystyle}

\newcommand{\beq}{\begin{equation}}
\newcommand{\eeq}{\end{equation}}
\newcommand{\beqan}{\begin{eqnarray*}}
\newcommand{\eeqan}{\end{eqnarray*}}

\newcommand{\openCase}  {\left\{ \begin{array}{@{\,}ll}}
\newcommand{\openCasell}{\left\{ \begin{array}{@{\,}ll}}
\newcommand{\openCasecl}{\left\{ \begin{array}{@{\,}cl}}
\newcommand{\openCaserl}{\left\{ \begin{array}{@{\,}rl}}
\newcommand{\openCaseTablell}{\left\{ \begin{array}{@{}ll}}
\newcommand{\openCaseTablecl}{\left\{ \begin{array}{@{}cl}}
\newcommand{\openCaseTablerl}{\left\{ \begin{array}{@{}rl}}
\newcommand{\closeCase} {\end{array} \right.}

\def\half{{\textstyle\frac{1}{2}}}

\DeclareMathOperator*{\argmin}{arg\,min}
\DeclareMathOperator*{\argmax}{arg\,max}

\DeclareMathOperator{\sign}{sgn}

\newcommand{\iP}[1]{\mathrm{P}({#1})} 
\newcommand{\E}[1]{\mathrm{E}\!\left[\,{#1}\,\right]}   
\newcommand{\varc}[2]{\mathrm{var}_{#2}\!\left({#1}\right)}   
\newcommand{\normal}[2]{{\mathcal{N}(#1,#2)}}  
\newcommand{\uniform}[2]{{\mathcal{U}(#1,#2)}}  

\def\muw{\mu_{\rm f}}

\def\E{\mathbb{E}}
\def\Ehist{E_{\rm hist}}

\def\muhatAlpha{\widehat{\mu}_\alpha}
\def\muhatDML{\widehat{\mu}_{\rm DML}}
\def\muhatGGML{\widehat{\mu}_{\rm GGML}}
\def\muhatQ{\widehat{\mu}_{\rm Q}}
\def\muhatQML{\widehat{\mu}_{\rm QML}}
\def\muhatMean{\widehat{\mu}_{\rm mean}}
\def\muhatMid{\widehat{\mu}_{\rm mid}}
\def\muhatNB{\widehat{\mu}_{\rm NB}}
\def\muhatNL{\widehat{\mu}_{\rm NL}}
\def\muV{\mu_{V}}
\def\muW{\mu_{W}}
\def\muX{\mu_{X}}
\def\muZ{\mu_{Z}}
\def\nvargg{\text{NVar(GGML)}}
\def\phat{\widehat{p}_{\Vtilde}}
\def\Psim{\Psi(m,x)}
\def\Psimu{\Psi(m,\muX)}
\def\sigD{\sigmaZ/\Delta}
\def\sigmaV{\sigma_V}
\def\sigmaW{\sigma_W}

\def\sigmaZ{\sigma_Z}
\def\Um{\mathcal{U}}

\def\Vtilde{\widetilde{V}}
\def\xQ{\widehat{\mu}_{\rm Q}}

\def\Xhist{X_{\rm hist}}

\begin{document}
\title{Estimation from Quantized Gaussian Measurements: When and How to Use Dither}

\author{Joshua~Rapp, Robin~M.~A.~Dawson, and~Vivek~K~Goyal%
\thanks{This work was supported in part by a Draper Fellowship under Navy contract N00030-16-C-0014,
  the U.S. National Science Foundation under Grant 1422034, and
  the DARPA REVEAL program under Contract No.\ HR0011-16-C-0030.
  This work was presented in part at the IEEE International Conference on Image Processing, Athens, Greece, October 2018.}%
\thanks{J. Rapp and V. K. Goyal are with the Department
of Electrical and Computer Engineering, Boston University, Boston,
MA, 02215 USA (e-mail: jrapp@bu.edu; v.goyal@ieee.org).}%
\thanks{J. Rapp and R. Dawson are with the Charles Stark Draper Laboratory, Cambridge, MA, 02139 USA.}%
}

\maketitle

\begin{abstract}
Subtractive dither is a powerful method for removing the signal dependence of quantization noise for coarsely-quantized signals.
However, estimation from dithered measurements often naively applies the sample mean or midrange, even when the total noise is not well described with a Gaussian or uniform distribution.
We show that the generalized Gaussian distribution 
approximately describes 
subtractively-dithered, quantized samples of a Gaussian 
signal. 
Furthermore, a generalized Gaussian fit leads to simple estimators based on order statistics that match the performance of more complicated maximum likelihood estimators requiring iterative solvers.
The order statistics-based estimators outperform both the sample mean and midrange for nontrivial sums of Gaussian and uniform noise.
Additional analysis of the generalized Gaussian approximation yields rules of thumb for determining when and how to apply dither to quantized measurements.
Specifically, we find subtractive dither to be beneficial when the ratio between the Gaussian standard deviation and quantization interval length is roughly less than 1/3.
If that ratio is also greater than 0.822/$K^{0.930}$ for the number of measurements $K>20$, we present estimators more efficient than the midrange.
\end{abstract}


\maketitle

\section{Introduction}
Estimation of the mean of a Gaussian distribution from
independent and identically distributed (i.i.d.)
samples is
a canonical problem in statistics, yet it has
important subtleties
when the samples are quantized.
Without quantization, the sample mean is an unbiased, efficient, and consistent estimator.
With uniformly quantized samples,
the situation is immediately more complicated:
The sample mean is an unbiased estimate only when the 
true mean falls on the quantizer's reproduction grid
or asymptotically in the limit of fine quantization~\cite{Gjeddebaek:59};
and
in the opposite extreme of very coarse quantization,
all the samples are identical, so the
estimates do not even improve
with increasing
numbers of data samples.

The use of subtractive dither changes the situation substantially.
The sample mean is then an unbiased and
consistent estimator---like in the unquantized case---but
it may be arbitrarily far from minimizing the
mean-squared error (MSE)\@.
For example, when the population variance vanishes,
the sample mean estimator has MSE inversely proportional
to the number of samples, whereas the 
MSE achieved by the midrange estimator 
is inversely proportional to the \emph{square} of the number of samples~\cite{Fisher1922}.

In this paper, we develop estimators for cases where the quantization is neither extremely fine nor extremely coarse.
The motivation for this work stemmed from a series of experiments performed by the authors and colleagues with single-photon lidar. 
In~\cite{Shin2016a}, temporally spreading a narrow laser pulse, equivalent to adding non-subtractive Gaussian dither, was found to reduce the effects of the detector's coarse temporal resolution on ranging accuracy.
Later work on a similar system showed that implementing subtractive dither could likewise reduce the effects of coarse quantization~\cite{Rapp2018a,Rapp2018b}.
Our aim was then to compare the two approaches by determining performance limits, optimal estimators, and when one method might be preferable over the other.
The estimators we develop in this work
are based on a generalized Gaussian (GG) approximation for the combination of sample variation
and quantization noise, which the authors first proposed in~\cite{Rapp2018b}.
While the benefit of the GG approximation did not yield improved results for the lidar data due to model mismatch, our framework is valid for a more general set of problems in which quantization of a Gaussian scalar signal occurs.
We propose a number of estimators for additive GG noise and find a clear computational advantage with negligible loss in accuracy for simple estimators based on order statistics.

\subsection{Main Contributions}
This paper makes the following contributions:
\subsubsection{Estimation Efficiency}
We demonstrate the inefficiency of the mean and midrange estimators for subtractively-dithered measurements of a Gaussian signal by deriving the maximum likelihood estimator and Cram{\'e}r-Rao bound.

\subsubsection{Generalized Gaussian Approximation}
We expand upon the generalized Gaussian approximation introduced in~\cite{Rapp2018b} for the sum of Gaussian and uniform random variables that arises from subtractively dithered quantization of a Gaussian signal, using the approximation to determine three distinct regimes of estimator behavior.

\subsubsection{Estimator Proposal}
We consider a family of location estimators based on the GG approximation, in particular linear combinations of the measurement order statistics. 
We introduce a version of the trimmed mean estimator with trimming determined by the GG approximation that is simple, computationally efficient, and performs as well as the ML estimator.
Monte Carlo estimator comparisons are shown versus the number of measurements $K$ and versus $\sigD$, the ratio of the Gaussian standard deviation to the quantization bin size.

\subsubsection{Rules of Thumb}
We determine several key rules of thumb for deciding when and how to use subtractive dither. 
For instance, we find dither is not beneficial roughly for $\sigD > 1/3$; below this value, however, applying subtractive dither and a GG-based estimator lowers the MSE.
Moreover, if the quantization is coarser than $\sigD=0.822/K^{0.930}$ and $K>20$, then the midrange is a good estimator.

\subsection{Outline}
This paper is organized as follows.
Section~\ref{sec:formulation} sets up the problem of measuring a Gaussian signal with a
subtractively-dithered quantizer and explores the fact that the mean and midrange are inefficient estimators.
Section~\ref{sec:GGmodel} motivates the use of the generalized Gaussian distribution and estimators based on order statistics.
Section~\ref{sec:implement} discusses several estimator implementations for our noise model.
Section~\ref{sec:regimes} introduces mean-squared error expressions as a guide to better understanding the results of numerical simulation presented in Section~\ref{sec:results}, which tests several estimators and compares the use of quantized data with and without dithering.
Finally, Section~\ref{sec:conclusion} presents our conclusions regarding which estimators to use and when to apply dither.
\section{Formulation, Background, and Motivation}\label{sec:formulation}
\subsection{Quantized Measurement}
We begin by presenting and expanding upon the signal acquisition model introduced in~\cite{Rapp2018b}.
Suppose we have an unknown constant signal $\muX$ corrupted by additive, zero-mean Gaussian noise $Z\sim \normal{0}{\sigmaZ}$.
Then estimation of $\muX$ from $K$ independent samples 
\begin{equation*}
    X_i = \muX + Z_i, \qquad i = 1,2,\dots,K,
\end{equation*}
is straightforward, as the sample mean
$\overbar{\mu}= (1/K) \sum_{i=1}^{K} X_i$
can easily be shown to be an efficient
estimator of the mean of a Gaussian distribution.
However, all measurement instruments perform some quantization.
For instance, consider a uniform midtread quantizer $q(\cdot)$ with bin size $\Delta$ applied to $X_i$ 
when $\sigmaZ \ll \Delta$.
Except when $\muX$ is close to a quantizer threshold, it will
be the case that $U_i = q(X_i)$ is identical for all $i$, so that the  
``quantized-sample mean'' given as 
\begin{equation}
    \xQ = \frac{1}{K}\sum_{i=1}^{K}U_i,
\end{equation}
is no more informative an estimate of $\muX$ than any single measurement.
For $\sigmaZ$ not too small compared to $\Delta$, estimation error can be reduced by properly accounting for the quantization and the underlying distribution, e.g., via the maximum likelihood estimator for quantized samples of a Gaussian signal~\cite{Vardeman2005,Moschitta2015}:
\begin{align}\label{eq:QML}
\muhatQML = \argmax_{\muX} \sum_{i=1}^K \log \Bigg [&\Phi \left (\frac{u_i-\muX+\frac{\Delta}{2}}{\sigmaZ}\right) -  \nonumber \\ 
    &\Phi \left (\frac{u_i-\muX-\frac{\Delta}{2}}{\sigmaZ}\right) \Bigg ],
\end{align}
where $\Phi(\cdot)$ is the cumulative distribution function (CDF) of a standard normal random variable.
Still, $\muhatQML$ is no more accurate than $\muhatQ$ when all of the samples have the same value.
Because of the coarse quantization mapping every value in $[j\Delta-\Frac{\Delta}{2},j\Delta+\Frac{\Delta}{2}]$ to $j\Delta$ for $j\in \mathbb{Z}$, the resolution of an estimate $\widehat{\mu}_X$ is limited by the bin size~$\Delta$ and the quantization error is signal-dependent.

Statisticians have long recognized that working with rounded data
is not the same as working with underlying continuous-valued data.
Let $\Xhist$ be the continuous random variable with
density constant on intervals $((j-\half)\Delta,\,(j+\half)\Delta)$
with $\iP{\Xhist \in ((j-\half)\Delta,\,(j+\half)\Delta)} = 
\iP{U = k\Delta}$, for all $j \in \Z$.
Because of the piecewise-constant form,
$\Xhist$ is said to have a \emph{histogram density}~\cite{Vardeman:05}.
The widely known \emph{Sheppard's corrections}
introduced in~\cite{Sheppard:1897,Sheppard:1898}
relate the moments of $U$ and the moments of $\Xhist$~\cite{Kendall:38}.
From the construction of $\Xhist$, it is immediate
that these corrections are zero for odd moments.
See~\cite{Heitjan:89} for a review of Sheppard's corrections
and~\cite{BaiZZH:09} for
results for autoregressive and moving average processes
and more recent references.

The moments of $\Xhist$ being close to the moments
of $X$ depends on continuity arguments and $\Delta$ being small.
In contrast, our interest here is in situations where the
quantization is coarse relative to the desired precision
in estimating $\muX$. 
Quantization may be coarse because of limitation of instruments,
such as the fundamental trade-offs in analog-to-digital converters~\cite{Walden:99}
or
the time resolution in
time-correlated single photon counting~\cite{OConnor1984},
which may be coarse relative to the resolution desired for
time-of-flight ranging.

When quantizing $\Xhist$,
the quantization error $\Ehist = q(\Xhist) - \Xhist$
is uniformly distributed on $[-\Delta/2,\,\Delta/2]$ and independent of $\Xhist$.
In general, however, quantization error being uniformly distributed and independent
of the input does not extend to the quantization of $X$;
approximating quantization error as such---without regard to whether the
input has a histogram density---is often called the
``additive-noise model,''
``quantization-noise model,'' or
``white-noise model.''
A substantial literature is devoted to understanding the
validity of this approximation,
e.g.~\cite{Widrow1961,Sripad1977,ClaasenJ:81,ViswanathanZ:01,MarcoN:05}.

One approach to improving the precision of estimates from quantization-limited measurements is the use of dither, a small signal introduced before the discretization to produce enough variation in the input such that it spans multiple quantization levels.
By combining multiple dithered measurements, estimates can achieve resolution below the least-significant bit and the result may also have desirable statistical and perceptual properties, such as whitened noise.
Early applications empirically demonstrating the benefits of dither include
control systems~\cite{maccoll1945fundamental,Ishikawa1960},
image display~\cite{Goodall1951,Roberts1962},
and audio~\cite{Jayant1972}, with numerous contributions to the statistical theory developed in~\cite{GGFurman1959,Jaffe1959,Widrow1961,Schuchman1964,Sripad1977,Gray1993,Wannamaker2000}, among others.
More recent work has focused on varying the quantizer thresholds primarily for 1-bit measurements in wireless sensing networks, including~\cite{Papadopoulos2001,Rousseau2003,Ribeiro2006,Dabeer2006,Fang2008,Farias2014,Gianelli2016}.
While these works consider various methods for optimizing or adapting thresholds, they are restricted to considering only nonsubtractively-dithered quantization.

\subsection{Subtractively-Dithered Quantization}
If it is possible to know the dither signal exactly, \emph{subtractively-dithered quantization} with the proper dither signal makes the quantization error uniformly distributed and independent of the input.
Define the dither signal $D_i,~i=1,\dots,K$ as a sequence of i.i.d. random variables, independent of the noisy quantizer input $X_i$.
The output of a subtractively-dithered quantizer is
\begin{equation}\label{eq:model}
    Y_i = q(X_i+D_i)-D_i,
\end{equation}
with the quantization error defined as
\begin{equation}
    W_i = Y_i-X_i.
\end{equation}
Define the characteristic function of the dither signal probability density function (PDF) as
\begin{equation}
    M_D(ju) = \E[e^{juD}].
\end{equation}
Then Schuchman's condition~\cite{Schuchman1964} is the property of the dither PDF that
\begin{equation}\label{eq:schuchman}
    M_D \left(j\Frac{2\pi\ell}{\Delta}\right) =  0,~\ell \in \Z \setminus 0.
\end{equation}
As long as the quantizer has a sufficient number of levels so that it does not overload, by~\cite{Sripad1977,Gray1993} the Schuchman condition is necessary and sufficient for $X_i$ to be independent of $W_j$ for all $i, j$, with 
i.i.d. $W_i\sim\Um[-\Delta/2,\Delta/2]$.
Subtractive dither often uses a uniform dither signal with $D\sim$ $\Um[-\Delta/2,\Delta/2]$ because its characteristic function
\begin{equation*}
    M_D(ju) = \frac{\sin(u\Frac{\Delta}{2})}{u\Frac{\Delta}{2}}
\end{equation*}
meets Schuchman's condition~\eqref{eq:schuchman}.

The rest of this paper considers only when Schuchman's condition is met, with an i.i.d. input signal of the form $X_i = \muX+Z_i,~Z\sim\mathcal{N}(0,\sigmaZ^2)$, an i.i.d. dither signal $D_i\sim$ $\Um[-\Delta/2,\Delta/2]$ independent of the input signal, and a non-overloading uniform quantizer.
Then the dithered measurements take the form
\begin{equation}
    Y_i = \muX + Z_i + W_i,
\end{equation}
and the problem of estimating $\muX$ simply becomes one of mitigating independent additive noise.
The sum of the Gaussian and uniform terms can be combined into a single total noise term to obtain
\begin{equation}
Y_i = \muX + V_i,
\end{equation}
where $V_i = Z_i + W_i$ are i.i.d.
Then the means and variances simply add so that $\muV=0$ and $\sigmaV^2 = \sigmaZ^2+\Delta^2/12$.

For convenient shorthand, we refer to measurements from a quantizer without dither as ``quantized'' and measurements from a subtractively-dithered quantizer as ``dithered.''
The usual approach to estimating $\muX$ from $K$ dithered measurements
$Y_i$, $i = 1,\,2,\,\ldots,\,K$, is via the sample mean
\begin{equation}
\muhatMean = \frac{1}{K}\sum_{i=1}^K Y_i.   
\end{equation}
The MSE of the sample mean is  
\begin{equation}\label{eq:mse_mean}
    \text{MSE(mean)} = \sigmaV^2/K,
\end{equation}
which is $O(K^{-1})$.
Although using the sample mean is logical when $\sigmaZ \gg \Delta$ so that the contribution of the uniform noise component is negligible,
the sample mean is not in general an efficient estimator.
For example, in an alternative case of $\sigmaZ = 0$,
a maximum likelihood (ML) estimator is the midrange\footnote{Any statistic in $[Y_{(n)}-\Frac{\Delta}{2},Y_{(1)}+\Frac{\Delta}{2}]$ is an ML estimator for the mean of a uniform distribution with known variance~\cite[p. 282]{Mood1974}. 
The midrange is commonly used because it is unbiased and the minimum-variance estimator among linear functions of order statistics~\cite{Lloyd1952}.
However, no uniformly minimum-variance unbiased estimator 
exists~\cite[p. 331]{Mood1974}.}
\begin{equation}
    \muhatMid = \frac{1}{2}\left(Y_{(1)}+Y_{(K)}\right),
\end{equation}
where $Y_{(1)}\leq Y_{(2)}\leq\dots\leq Y_{(K)}$ are the order statistics of the $K$  measured samples.
Whereas the MSE of the sample mean for $\sigmaZ = 0$ is $\Delta^2/(12K)$,
the MSE of the midrange is

\begin{equation}\label{eq:mse_mid}
    \text{MSE(mid)} = \Frac{\Delta^2}{[2(K+1)(K+2)]},
\end{equation}
which is $O(K^{-2})$ and hence better than the sample mean
by an unbounded factor~\cite{Lloyd1952}.
Nevertheless, the midrange is not a good
estimator
in the general case of $\sigmaZ > 0$,
as it relies on the finite support of the uniform distribution.
If instead
$\sigmaZ$ is
much larger than $\Delta$, 
rendering the uniform component negligible, then
the MSE of the midrange would only improve as $O(1/\log(K))$~\cite{Broffitt1974}.
As others have noted for quantization of a Gaussian signal without dither~\cite{Vardeman2005,Moschitta2015}, the key figure of merit for determining estimator performance is then $\sigD$, a measure of the relative sizes of the noise components.
We observe that normalizing the MSE by $\Delta^2$ removes the separate dependence on $\sigmaZ$ and $\Delta$, resulting in
\begin{equation}\label{eq:nmse_mean}
    \text{NMSE(mean)} = [(\sigD)^2+1/12]/K,
\end{equation}
and
\begin{equation}\label{eq:nmse_mid}
    \text{NMSE(mid)} = \Frac{1}{[2(K+1)(K+2)]}.
\end{equation}

Except in trivial cases ($\sigmaZ\gg\Delta$ or  $\sigmaZ\ll\Delta$), $V$ has neither Gaussian nor uniform distribution, so the conventional mean and midrange estimators are expected to be suboptimal.
Furthermore, existing nonlinear processing schemes for dithered measurements do not adapt to best suit the noise statistics~\cite{Carbone1996}.
A first approach to finding a better estimator for arbitrary $\sigD$ is to derive the maximum likelihood (ML) estimator for the dithered noise model.
From the definitions of the random variables, the PDF of $W$ is 
\begin{equation*}
    f_W(w) = \begin{cases} \Frac{1}{\Delta}, & w\in [-\Delta/2,\Delta/2] \\
    0, & \text{otherwise},
    \end{cases}
\end{equation*}
and the PDF of $Z$ is $f_Z(z) = \phi(\Frac{z}{\sigmaZ})/\sigmaZ$, where $\phi(x)$ is the standard normal PDF\@.
Since the total noise is the sum of independent noise terms, the PDF of the samples is given by the convolution
\begin{align}\label{eq:true_pdf}
f_V(v) &= f_Z(z)*f_W(w) \nonumber \\
    &= \frac{1}{\Delta}\int_{-\frac{\Delta}{2}}^{\frac{\Delta}{2}}f_Z(v-\tau)d\tau   \nonumber \\
    &= \frac{1}{\Delta} \left [\Phi \left (\frac{v+\frac{\Delta}{2}}{\sigmaZ}\right) - \Phi \left (\frac{v-\frac{\Delta}{2}}{\sigmaZ}\right) \right ].
\end{align}
For i.i.d.\ samples from a dithered quantizer, the likelihood function is then 
\begin{align}
\mathcal{L} &\left(\{y_i\}_{i=1}^K;\muX \right) = \prod_{i=1}^K f_V(y_i-\muX) \nonumber \\
    &= \prod_{i=1}^K \frac{1}{\Delta} \left [\Phi\!\left (\frac{y_i-\muX+\frac{\Delta}{2}}{\sigmaZ}\right) - \Phi\!\left (\frac{y_i-\muX-\frac{\Delta}{2}}{\sigmaZ}\right)\right ].
\end{align}
From the log-likelihood, the dithered-sample ML estimator of $\muX$ is
\begin{align}\label{eq:DML}
\muhatDML = \argmax_{\muX} \sum_{i=1}^K \log \Bigg [&\Phi \left (\frac{y_i-\muX+\frac{\Delta}{2}}{\sigmaZ}\right) -  \nonumber \\ 
    &\Phi \left (\frac{y_i-\muX-\frac{\Delta}{2}}{\sigmaZ}\right) \Bigg ].
\end{align}
The ML estimator is notably identical to \eqref{eq:QML}, except the dithered measurements are not discrete-valued as are the samples used for $\muhatQML$.

To determine the efficiency of the mean, midrange, and DML estimators, we derive the Cram{\'e}r-Rao bound (CRB), which is a limit on the MSE that an unbiased estimator can achieve~\cite[Chapter 4.2.2]{VanTrees2013}.
The normalized CRB is derived in Appendix~\ref{sec:crb} for one dithered measurement to be 
\begin{align}\label{eq:ncrb}
    \text{NCRB}(\muX) 
    &= \frac{(\sigD)^2}{\ds \int  \frac{\left[\phi\left(\frac{u-1/2}{\sigD}\right)-\phi\left(\frac{u+1/2}{\sigD}\right) \right]^2}{\Phi\left(\frac{u+1/2}{\sigD}\right)-\Phi\left(\frac{u-1/2}{\sigD}\right)} du},
\end{align}
which can be evaluated via numerical integration.
Note that the uniform PDF does not meet the regularity condition required for the CRB to apply, so~\eqref{eq:ncrb} is not expected to be meaningful for $\sigD=0$.

\begin{figure}
    \centering
    \includegraphics[trim={2cm 1.3cm 2cm 1.3cm},clip,width=0.7\linewidth]{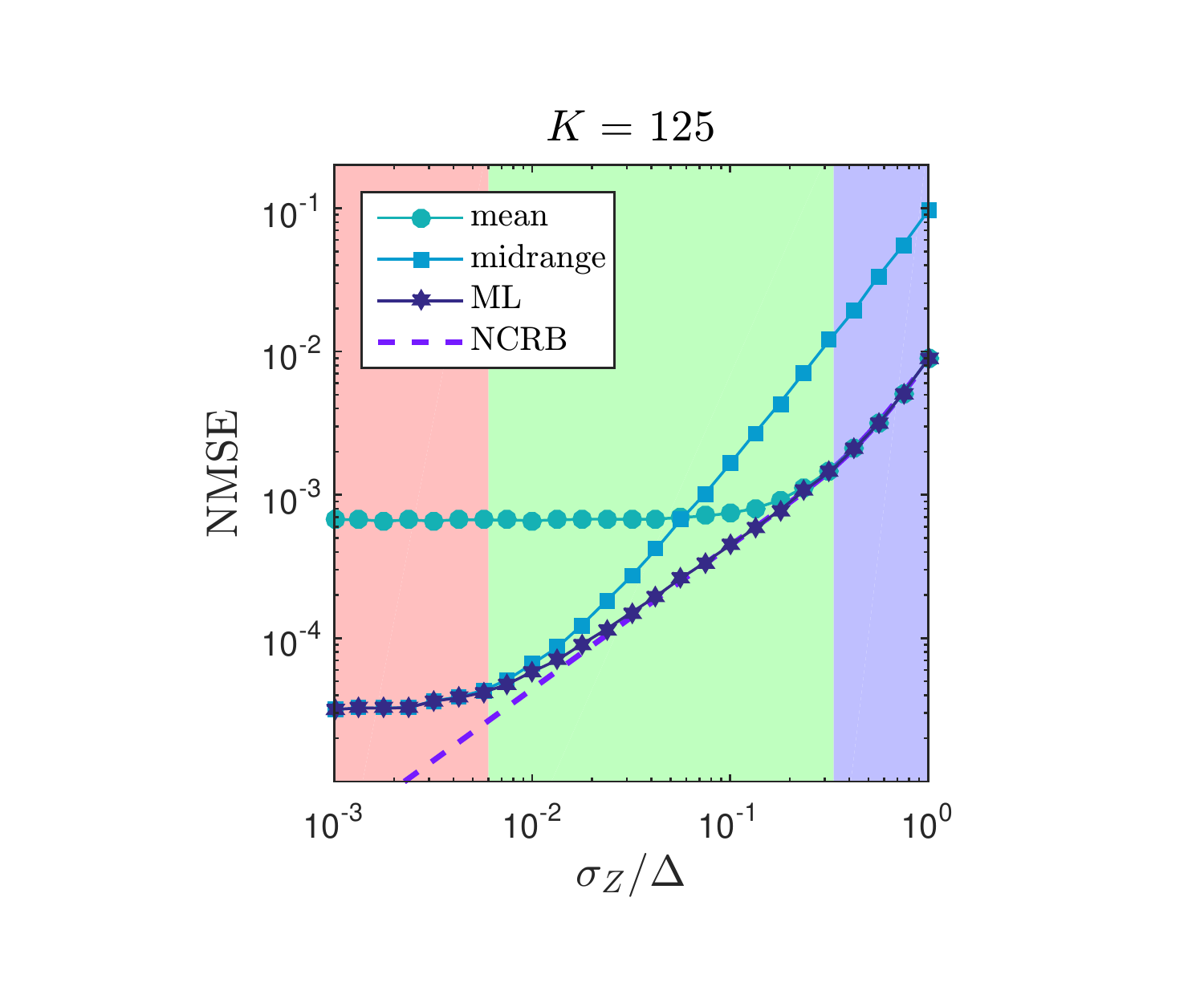}
    \caption{
    Applied to subtractively-dithered measurements, the midrange is approximately optimal only for small $\sigD$ (Regime~I, red), and the sample mean is approximately optimal only for large $\sigD$ (Regime~III, blue).
    For confirmation, in Regime~I the midrange approaches NMSE $= 1/[2(K+1)(K+2)] \approx 3.12\times10^{-5}$ and the mean approaches NMSE $= (1/12)/K \approx 6.67\times10^{-4}$, and in Regime~III the mean approaches NMSE $= (\sigD)^2/K$, which is $8.00\times10^{-3}$ at $\sigD=1$.
    We seek an estimator simpler than the dithered-sample maximum likelihood that performs at least as well as the mean and midrange for intermediate values of $\sigD$ (Regime~II, green).
}
    \label{fig:example}
\end{figure}

Fig.~\ref{fig:example} illustrates the suboptimality of the mean and midrange estimators compared to $\muhatDML$ for intermediate values of $\sigD$.
In a Monte Carlo simulation with $T = 20000$ trials, $K=125$ measurements were generated according to~\eqref{eq:model}, where both $\muX$ and $D$ were selected uniformly at random over $[-\Delta/2,\Delta/2]$.
Computing the normalized MSE of the $\muhatMean$, $\muhatMid$, and $\muhatDML$ estimates as
\begin{equation}
    \text{NMSE}(\widehat{\mu}_X) = \frac{1}{T}\sum_{t=1}^{T} \left ( \frac{\muX-\widehat{\mu}_X}{\Delta}\right )^2
\end{equation}
reveals how the performance of each estimator changes as a function of $\sigD$.

Fig.~\ref{fig:example} highlights three distinct regimes of estimator behavior.
In Regime~I (red), the Gaussian noise component is negligible, so the ML estimator and the midrange are nearly identical and outperform the mean.
In Regime~III (blue), the uniform noise component is negligible, so the ML estimator and the mean are nearly identical and outperform the midrange.
In Regime~II (green), neither the uniform nor the Gaussian component dominates, and the DML estimator performs significantly better than both the mean and midrange.
Still, $\muhatDML$ does not achieve the CRB for small $\sigD$, indicating that an efficient estimator of $\muX$ does not exist; however, 
all ML estimators are
asymptotically efficient in $K$~\cite{VanTrees2013}.

From the results in Fig.~\ref{fig:example}, it may seem obvious that $\muhatDML$ is a better choice than $\muhatMean$ or $\muhatMid$ for any value of $\sigD$.
However, $\muhatDML$ requires iterative solution, thus making it far more computationally complex than the mean and midrange.
In this work, one of our primary aims is to find a computationally simple estimator that can likewise outperform the mean and midrange in Regime~II\@.
We show that a generalized Gaussian approximation to the total noise of a dithered quantizer gives rise to order statistics-based estimators that approach the performance of $\muhatDML$.
In addition, we compare their results to those from quantized measurements without dither, leading to design rules for when to use dither and which estimator to apply.

\subsection{Mixed Measurements of Vector Signals}
Whereas this paper is focused on the estimation of a single scalar value from multiple direct noisy measurements,
many other estimation problems involving quantized data have been studied extensively.
In particular,
interest in linear inverse problems---both undersampled and oversampled---has resulted in work on estimating vectors from quantized linearly mixed measurements,
with and without subtractive dither.

Quantized, mixed, noisy measurements of a vector $\textbf{x}$ can be represented as
  $\textbf{y} = Q(A\textbf{x} + \textbf{z})$,
where $A \in \C^{K \times N}$ is a linear operator,
$\textbf{z}$ is additive noise, and
$Q(\cdot)$ represents scalar quantization.
Theoretical results on how well $\textbf{x}$
can be estimated generally depend on
the structure of $A$,
such as being an oversampled inverse discrete Fourier transform---thus modeling oversampled analog-to-digital conversion (OADC)---or being large and random.

For noiseless ($\textbf{z} = 0$) OADC,
$O(K^{-2})$ upper- and lower-bounds on MSE using deterministic analyses~\cite{ThaoV:94,ThaoV:96} are reminiscent of \eqref{eq:mse_mid} and similarly rooted in
quantized values providing hard constraints on $\textbf{x}$.
For general $A$,
such constraints can be expressed with a linear program~\cite{GoyalVT:98},
and introduction of subtractive dither makes the
optimal $O(K^{-2})$ rate provably achievable by
a very simple algorithm~\cite{RanganG:01}.
The compressive case of $K < N$ is addressed,
for example, in~\cite{JacquesHF:11}.

For Gaussian $\textbf{z}$,
\cite{Zymnis2010} provides a method applicable
with nonuniform quantization to provide an
$\ell_1$-regularized estimate of $\textbf{x}$.
More general priors and quantizers
(potentially non-regular as well as nonuniform)
can be incorporated in the method of~\cite{KamilovGR:12}.
Like these earlier works,
this paper also addresses the case of Gaussian $\textbf{z}$,
but $A$ being a $K \times 1$ matrix of $1$s
makes it qualitatively different,
in part because the estimate of a single scalar from
$K \gg 1$ measurements need not be regularized.

\section{Generalized Gaussian Approximation and Estimation}\label{sec:GGmodel}

In order to find a simple estimator for Regime II, we begin by examining the other two regimes and the simple forms of the ML estimator there.
We notice that the uniform and Gaussian noise distributions in Regimes I and III are special cases of the generalized Gaussian distribution (GGD), which has PDF~\cite{varanasi1989parametric}
\begin{equation}\label{eq:ggpdf}
    f_{\Vtilde}(v;\mu,\sigma,p) = \frac{1}{2\Gamma(1+1/p)A(p)}\exp \left \{-\left ( \frac{|v-\mu|}{A(p)}\right )^p  \right \},
\end{equation}
where
$A(p) = \sqrt{\Frac{\sigma^2 \Gamma(1/p)}{\Gamma(3/p)}}$
and $\Gamma(\cdot)$ is the Gamma function.
In addition to mean and variance parameters $\mu$ and $\sigma^2$, the GG density has a third parameter $p$ that controls the exponential decay of its tails.
When $p=2$ or $p\rightarrow \infty$, the GGD simplifies to the Gaussian or uniform distributions, respectively.
Another special case of the GGD is the Laplace distribution for $p=1$.

For each of the special cases, we further notice that the ML estimator (median, mean, and midrange for $p= 1,~2,~\infty$) is a linear combination of order statistics.
When $p=1$, only the middle order statistic has nonzero weight, whereas the reverse is true for $p\rightarrow \infty$, with all weight on the two extreme samples.
For $p=2$, all of the order statistics are equally weighted.
With these two observations in mind, we hypothesize that, if there is a value of $p$ that approximates intermediate combinations of uniform and Gaussian noise, then there may be a corresponding order statistics-based estimator that approaches the performance of $\muhatDML$.

\subsection{Approximation}
\begin{figure}
    \centering
    \includegraphics[width=0.9\linewidth]{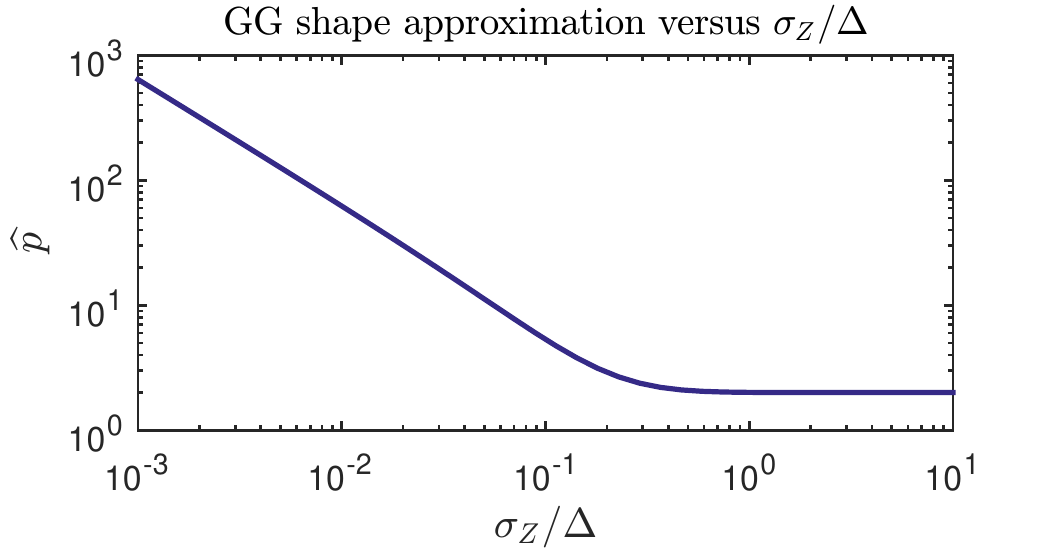}
    \caption{The value of $\phat$ goes to infinity as $\sigD$ decreases and converges to $\phat=2$ as $\sigD$ increases, with convergence beginning around $\sigD=1/3$ matching the anticipated behavior.}
    \label{fig:p_v_sd}
\end{figure}

For our stated purpose, it would be ideal if
proper selection of $p$ exactly represented nontrivial sums of  Gaussian and uniform terms.
Unfortunately the sum of any two independent GG random variables (GGRVs) is another GGRV only when $p=2$ for each addend%
\footnote{The limiting distribution of the sum of i.i.d. GGRVs is Gaussian by the Central Limit Theorem~\cite[Chapter 5.4.2]{Mood1974}, but the sum of any finite number of GGRVs will only be approximately Gaussian unless each term is Gaussian.}%
~\cite{Zhao2004}.
Nevertheless, the sum of independent GGRVs has many of the same properties as a GGRV, and can be well-approximated as a GGRV through a number of approximation methods.
A simple approach from~\cite{Soury2015} matches the mean, variance, and kurtosis of the GG approximation to the corresponding moments of the true noise distribution as follows.
Defining $\Vtilde$ as the GG approximation to $V=Z+W$, then
since the uniform and Gaussian noise components are independent random variables,
the mean and variance parameters of the GG noise approximation are simply given as $\mu_{\Vtilde} = \muW+\muZ$ and $\sigma_{\Vtilde}^2 = \sigmaZ^2+\sigmaW^2$.
To compute the shape parameter for the special case of uniform and Gaussian addends, the approximation of $p$ is then the unique solution to 
\begin{align}
    \frac{\Gamma(1/\phat)\Gamma(5/\phat)}{\Gamma(3/\phat)^2} 
    &= 3-\frac{6}{5}\frac{1}{\left [12\left (\frac{\sigmaZ}{\Delta}\right)^2+1 \right]^2}\label{eq:match_kurt}
\end{align}
(see derivation in Appendix~\ref{sec:kurtosis}).
We thus see that $\phat$ depends on $\sigD$, with the relationship plotted in Fig.~\ref{fig:p_v_sd}.
Solving~\eqref{eq:match_kurt} is fast, and the values of $\phat$ for a range of $\sigD$ values could be precomputed and stored in a table if necessary.
A rough approximation and good initial value for a solver is $\phat^{(0)} = \max\{2,\Delta/\sigmaZ\}$.

\begin{figure}
    \centering
    \includegraphics[width=\linewidth]{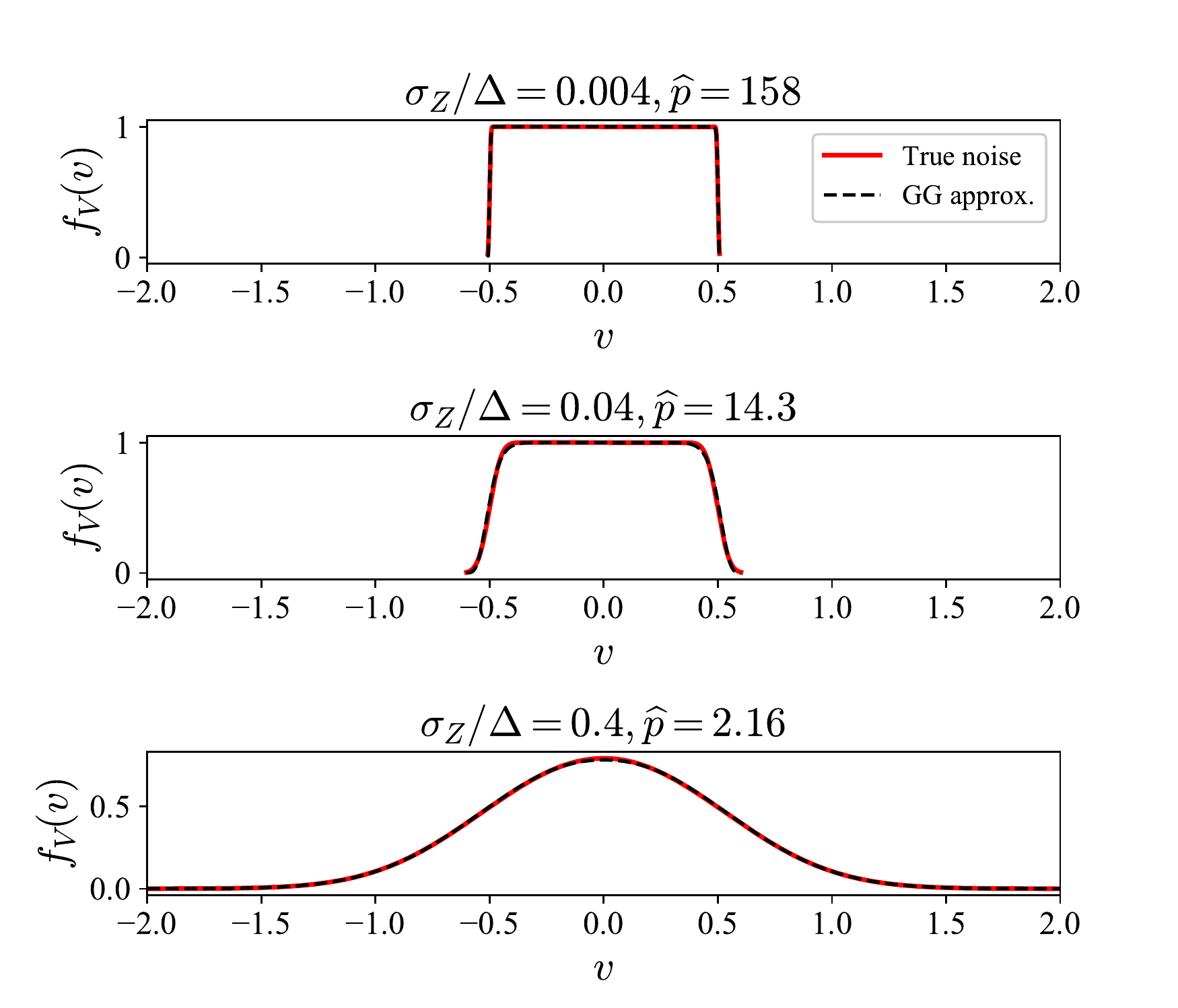}
    \caption{The three plots show the noise PDF calculated numerically from the true density~\eqref{eq:true_pdf} (solid red) and via the GG approximation~\eqref{eq:ggpdf} (dashed black).
    The close agreement suggests the GGD is a good approximation for the noise.
    }
    \label{fig:ggd_match}
\end{figure}

To verify the quality of the generalized Gaussian approximation to the output noise distribution using the kurtosis match, Fig.~\ref{fig:ggd_match} shows comparisons between the true density, computed numerically according to~\eqref{eq:true_pdf}, and its GG approximation from~\eqref{eq:ggpdf}.
We test $\sigD=$ 0.004, 0.04, and 0.4, maintaining $\Delta=$~1 for consistency.
For $\sigmaZ\ll\Delta$, the distribution is close to uniform, and at $\sigmaZ\approx\Delta$, the distribution is almost Gaussian.
In the intermediate regime, however, the distribution combines attributes of each component, with the flat top of the uniform distribution and exponential tails of the Gaussian distribution.
The GGD appears to be a good approximation of the true noise distribution, almost perfectly matching the shape behavior.

\subsection{Estimation}
\begin{figure}
    \centering
    \includegraphics[width=0.9\linewidth]{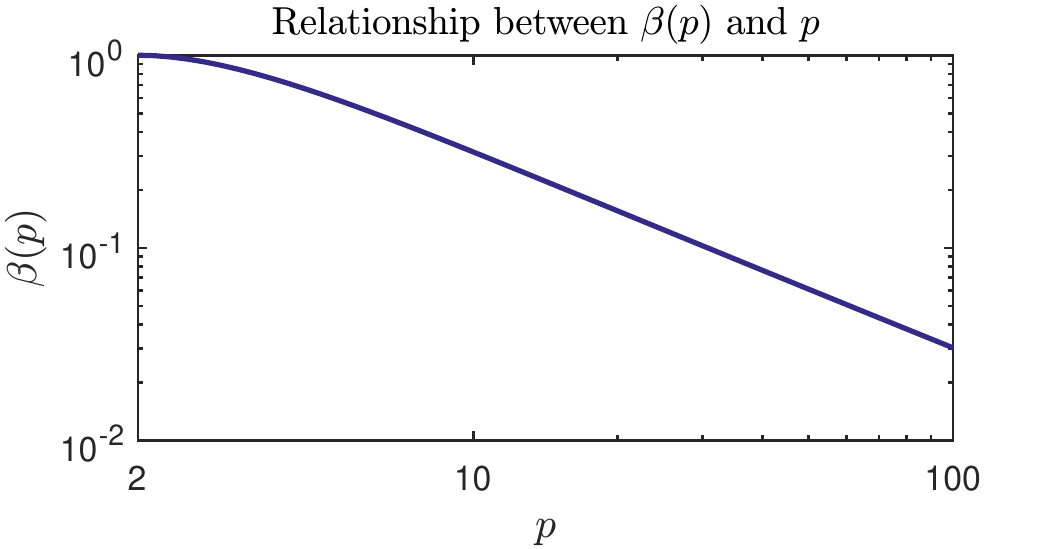}
    \caption{As $p$ increases beyond $p=2$, $\beta(p)$ becomes much less than 1, implying that MSE$(\muhatGGML)$ is much lower than MSE$(\muhatMean)$.}
    \label{fig:beta_v_p}
\end{figure}

For i.i.d. samples of a GG distribution, the likelihood function is 
\begin{equation}\label{eq:gglikelihood}
    \mathcal{L}(\{v_i\}_{i=1}^K;\mu,\sigma,p) = \prod_{i=1}^K f_{\Vtilde}(v;\mu,\sigma,p).
\end{equation}
By differentiating the log of~\eqref{eq:gglikelihood} with respect to $\mu$, the ML estimator $\muhatGGML$ for the mean of a GGRV is 
given in~\cite{varanasi1989parametric} as the solution to 
\begin{equation}\label{eq:GGML}
    \sum_{i=1}^K \sign(y_i-\muhatGGML) |y_i-\muhatGGML|^{p-1} = 0,
\end{equation}
and is shown to be asymptotically normal and efficient in $K$ for $p\geq2$, which is the regime of interest.
The asymptotic variance of $\muhatGGML$ normalized by $\Delta^2$ is given by
\begin{equation}\label{eq:asympGGML}
    \nvargg = \frac{\beta(p)[(\sigD)^2+1/12]}{K},
\end{equation}
where
\begin{equation}
    \beta(p) =  \frac{\Gamma^2(1/p)}{p^2 \Gamma \left (\frac{2p-1}{p}\right)\Gamma(3/p)}.
\end{equation}
We notice that~\eqref{eq:asympGGML}
decreases as $O(K^{-1})$, but the coefficient $\beta(p)$, which is plotted in Fig.~\ref{fig:beta_v_p}, is much less than 1 for large $p$, suggesting that $\muhatGGML$ should outperform $\muhatMean$ for $\phat > 2$.
Since the GGD closely approximates the total noise distribution, it would be ideal if $\muhatGGML$ reduced to a computationally simple estimator such as one based on order statistics for all $p$.
Unfortunately, the ML estimator does not generally have a closed-form expression, except in special cases such as $p=1,2,\infty$ (an explicit expression has also recently been derived for $p=4$ \cite{Beaulieu2013}), so iterative solution would again be necessary.

We have already observed that the ML estimators for $p=1,2,\infty$ all belong to a class of linear combinations of order statistics called L-estimates~\cite{arce2005nonlinear}, which are attractive because they have closed-form definitions of the form
\begin{equation}\label{eq:l-est}
    \widehat{\mu}_X = \sum_{i=1}^K a_iY_{(i)}.
\end{equation}
We thus consider how to obtain the coefficients $a_i$ for L-estimates that perform well for GG noise when $p$ is not one of the special cases.

An effective
L-estimate should weight the order statistics in accordance with the noise distribution.
Notable past approaches include that of Lloyd, who derived the minimum-variance unbiased estimator among linear functions of the order statistics in~\cite{Lloyd1952}.
This formulation is impractical, however, as it requires the correlations of the order statistics for a given distribution, which are often not known even for common special cases like the Gaussian distribution.
Bovik et al.~\cite{Bovik1982,Bovik1983} further specified the minimum variance unbiased L-estimate and then numerically computed results for several values of $p$ from samples of a GGD with $K=3$.

A number of approximations to Lloyd's formulation exist to more simply compute near-optimal coefficients for linear combinations of order statistics, including~\cite{Gupta1952,Blom1957}.
{\"O}ten and de Figueiredo introduced one such method using Taylor expansion approximations to get around the difficulties of knowing distributions of order statistics~\cite{Oten2003}.
This method does still require knowledge of the inverse CDF of the noise distribution, and while there is no closed form expression for the GGD, the necessary values can be pre-computed numerically.

Simpler L-estimates have much longer histories, with consideration of trimming extreme or middle order statistics at least as old as~\cite{Anonymous1821} (credited to Gergonne in~\cite{Stigler1976}), with the first known mathematical analysis by Daniell, who called such an estimate the ``discard-average''~\cite{Daniell1920,Stigler1973}.
The method now known as the $\alpha$-trimmed mean and popularized by Tukey~\cite{Tukey1962,Tukey1963} avoids extensive computation of the weights by trimming a fixed fraction $\alpha$
from the extremes of the order statistics.
Restrepo and Bovik defined a complementary $\alpha$-``outer'' trimmed-mean~\cite{Restrepo1988}, which 
retains a fraction $\alpha$ of the data by trimming
the middle order statistics and is suitable for distributions with short tails within the range from Gaussian to uniform distributions.
They tabulated several instances of the trimmed mean for GGDs with multiple combinations of $K$ and $p$.

Lastly, Beaulieu and Guo introduced an estimator specifically for the GGD but using nonlinear combinations of the order statistics~\cite{Beaulieu2012}.
The weighting of the order statistics depends on $p$ via a heuristically-justified function and is shown to perform almost identically to $\muhatGGML$. 
This estimator is unbiased and exactly matches the ML estimator for the special cases of $p = 2$ and $\infty$.

In the following section, we consider three of the most computationally-efficient order statistics-based estimators to use for the GG approximation: the nearly-best L-estimate $\muhatNB$ of~\cite{Oten2003}, the trimmed-mean estimator $\muhatAlpha$ modeled on~\cite{Restrepo1988}, and the non-linear estimator $\muhatNL$ of~\cite{Beaulieu2012}.
Each estimator takes the form of~\eqref{eq:l-est} with different computations of the coefficients $a_i$.
While $\muhatNL$ is specifically designed for use with GG noise, we modify the more general $\muhatNB$ and $\muhatAlpha$ to match the GG approximation.
For $\muhatNB$, we use the PDF and inverse CDF (computed numerically) of the GG approximation to determine the coefficients.
One could alternatively compute the coefficients for $\muhatNB$ directly from the true noise distribution in~\eqref{eq:true_pdf}; however, additional numerical evaluation would be required for the inverse CDF, which we eschew in our search for computationally efficient estimators.
There is no explicit distribution assumed by $\muhatAlpha$, but we propose a choice of the trimmed fraction $\alpha$ based on the estimated $\phat$ value to implicitly link the estimator to the GGD.

\section{Estimator Implementations}\label{sec:implement}
\subsection{ML Estimators}\label{sec:MLimplement}
An EM algorithm for obtaining the quantized-sample ML estimate $\muhatQML$ was introduced by Papadopoulos et al.~\cite[Appendix E]{Papadopoulos2001}:
\begin{align}\label{eq:EM_ML}
    \muhatQML^{(j+1)} &= \muhatQML^{(j)} + \frac{\sigmaZ}{K\sqrt{2\pi}}\sum_{i=1}^K m(u_i),
\end{align}
where
\begin{align}
    m(u_i) = \frac{\exp\!\left(-\frac{\left[u_i-\frac{\Delta}{2}-\muhatQML^{(j)}\right]^2}{2\sigmaZ^2} \right)-\exp\!\left(-\frac{\left[u_i+\frac{\Delta}{2}-\muhatQML^{(j)}\right]^2}{2\sigmaZ^2} \right)}{\Phi\!\left(\frac{u_i+\frac{\Delta}{2}-\muhatQML^{(j)}}{\sigmaZ} \right)-\Phi\!\left(\frac{u_i-\frac{\Delta}{2}-\muhatQML^{(j)}}{\sigmaZ} \right)}.
\end{align}
A good initialization is $\muhatQML^{(0)}=\muhatQ$, since the estimators are equal for $\sigmaZ = 0, \infty$. 
A similar algorithm, derived in~\cite{Zymnis2010} for quantized, linearly-mixed vector measurements, is equivalent to that in~\eqref{eq:EM_ML} for the special case of a repeated scalar input and no mixing (i.e., the mixing matrix is a column of 1s).
Since $\muhatDML$ has the same formulation as $\muhatQML$, the same algorithm also works for continuous-valued dithered measurements:
\begin{align}
    \muhatDML^{(j+1)} &= \muhatDML^{(j)} + \frac{\sigmaZ}{K\sqrt{2\pi}}\sum_{i=1}^K m(y_i).
\end{align}
We initialize with $\widehat{\mu}_{\rm DML}^{(0)}=\muhatMid$, since the midrange is known to be the ML estimator for $\sigmaZ=0$.
A solver for $\muhatGGML$ was likewise initialized with $\muhatGGML^{(0)} = \muhatMid$.

\subsection{Order Statistics-Based Estimators}
To evaluate the GG noise approximation and find the best non-iterative estimator, we compared the three simplest estimators based on the order statistics: the nearly-best L-estimate, the $\alpha$-outer mean, and the nonlinear combination from~\cite{Beaulieu2012}.
Since the GGD is symmetric, the coefficients of an unbiased order statistics-based estimator are defined symmetrically and only half must be uniquely computed.
It is thus useful to define $M = \lfloor K/2 \rfloor$ and $N = \lceil K/2 \rceil$ using the floor and ceiling functions, respectively.

To derive the nearly-best L-estimate of~\cite{Oten2003}
\begin{equation}
    \widehat{\mu}_{\rm NB} = \sum_{i=1}^K a^{\rm NB}_i Y_{(i)},
\end{equation}
we first compute
\begin{subequations}
\begin{align}
    b_1 &= f_V(c_1)[-2f_V(c_1)+f_V(c_2)], \\
    b_i &= f_V(c_i)[f_V(c_{i-1}-2f_V(c_1)+f_V(c_{i+1})], \\
        &  \qquad \qquad \qquad \qquad i=2,\dots,N-1, \nonumber \\
    b_{N} &= f_V(c_{N})[f_V(c_{{N}-1})-f_V(c_{N})],
\end{align}
\end{subequations}
where $c_i = F_V^{-1}(i/(K+1))$, and $F_V^{-1}$ is the inverse of the GG CDF\@.
From this, the weights are derived for $i=1,\dots,N$ as
\begin{equation}
a^{\rm NB}_i = a^{\rm NB}_{K-i+1} =  \openCaserl
      \Frac{b_i}{\left (2\sum_{i=1}^N b_i \right )}, & K~\text{even}; \\
      \Frac{b_i}{\left (b_N + 2\sum_{i=1}^{M} b_i \right )}, & K~\text{odd}.
   \closeCase    
\end{equation}
For the simulations in Python, the inverse CDF was numerically computed with the \texttt{stats.gennorm.ppf} GGD percentile function in \texttt{scipy}, as no closed-form expression exists.

For the $\alpha$-outer mean estimate 
\begin{equation}
    \widehat{\mu}_\alpha = \sum_{i=1}^K a^\alpha_i Y_{(i)},
\end{equation}
the order statistics' weights $a^\alpha_i$ are only given in~\cite{Restrepo1988} for a symmetric filter applied to an odd number of samples:
\begin{equation}
a^\alpha_i = a^\alpha_{K-i+1}=  \openCaserl
      \ds \frac{1}{K\alpha},
                  & i \leq \lfloor\half K\alpha \rfloor \\
      \ds \frac{\half K\alpha -\lfloor\half K\alpha \rfloor}{K\alpha},
                  & i = \lfloor\half K\alpha \rfloor + 1, \\
                  & \quad \alpha\in[0,\,1-1/K]; \\
      \ds \frac{K\alpha - 2\lfloor \half K\alpha \rfloor}{K\alpha},
                  & i = \lfloor\half K\alpha \rfloor + 1, \\
                  & \quad \alpha\in[1-1/K,\,1]; \\
      0, & \text{otherwise}.
   \closeCase
\end{equation}
Since an even number of measurements is also possible, we similarly define,
for all $\alpha \in [0,1]$,
\begin{equation}
a^\alpha_i = a^\alpha_{K-i+1}=  \openCaserl
      \ds \frac{1}{K\alpha},
                  & i \leq \lfloor \half K\alpha \rfloor; \\
      \ds \frac{\half K\alpha-\lfloor\half K\alpha \rfloor}{K\alpha},
                  & i = \lfloor \half K\alpha \rfloor + 1; \\
      0,          & \text{otherwise}.
   \closeCase
\end{equation}
Note that the outer mean is equivalent to $\muhatMean$ when $\alpha = 1$ and reduces to $\muhatMid$ for $\alpha= 0$.
To match the GGD behavior, we thus propose to define
$\alpha = \Frac{2}{\phat}$,
which yields the ML estimate for both $\phat=2$ and $\phat=\infty$.
Finally, the nonlinear estimator of~\cite{Beaulieu2012} is given as
\begin{equation}
    \widehat{\mu}_{\rm NL} = \sum_{i=1}^{K} a_i^{\rm NL}Y_{(i)}, 
\end{equation}
where the data-dependent coefficients are given for $i = 1,\dots,M$ by
\begin{equation}
    a_i^{\rm NL} = a^{\rm NL}_{K-i+1} = \frac{1}{2}\frac{[Y_{(K-i+1)}-Y_{(i)}]^{p-2}}{\sum_{j=1}^{M} [Y_{(K-j+1)}-Y_{(j)}]^{p-2}}.
\end{equation}
Note that if $K$ is odd, the median term ($i=N$) is ignored, as it would correspond to a numerator of zero.

\section{Dither Noise Regimes}\label{sec:regimes}
To better understand the dither noise behavior, we have previously described three regimes of the dither noise distribution, with Regimes I and III corresponding to approximately uniform and Gaussian noise, respectively.
We have furthermore proposed the GGD with $p\in(2,\infty)$ as an approximation for the noise distribution in Regime II\@.
However, the boundaries of these regions are imprecise, and we aim to more rigorously define them in this section.
We first define $\xi_1$ and $\xi_2$ as the values of the ratio $\sigD$ separating the regimes such that the noise distribution is approximately uniform for $\sigD<\xi_1$, GG for $\xi_1\leq\sigD<\xi_2$, and Gaussian for $\sigD\leq\xi_2$.
In each regime, we have an expression for the expected MSE or asymptotic variance of the ML estimator, so we use the intersection or approximate point of convergence of these expressions to define $\xi_1$ and $\xi_2$.

\subsection{Defining $\xi_1$}
\begin{figure}
    \centering
    \begin{subfigure}{0.49\linewidth}
        \centering
        \includegraphics[width=\linewidth]{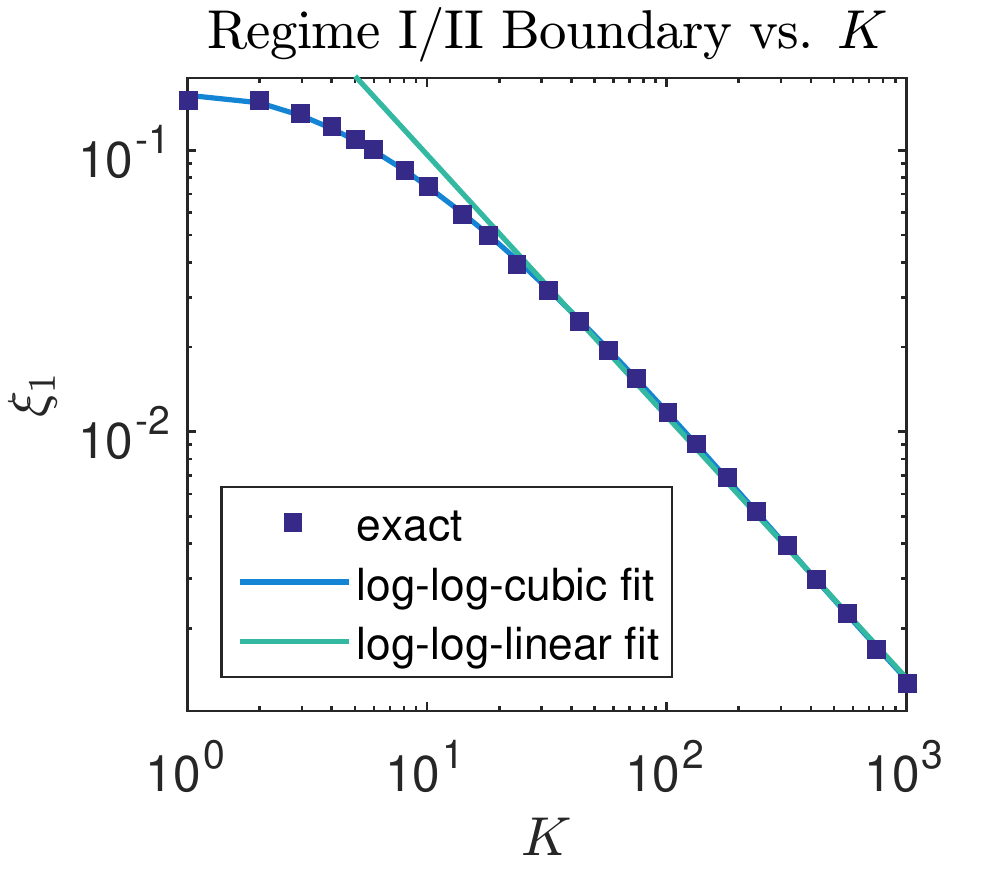}
        \caption{}
        \label{fig:xi1}
    \end{subfigure}
    \begin{subfigure}{0.49\linewidth}
        \centering
        \includegraphics[width=\linewidth]{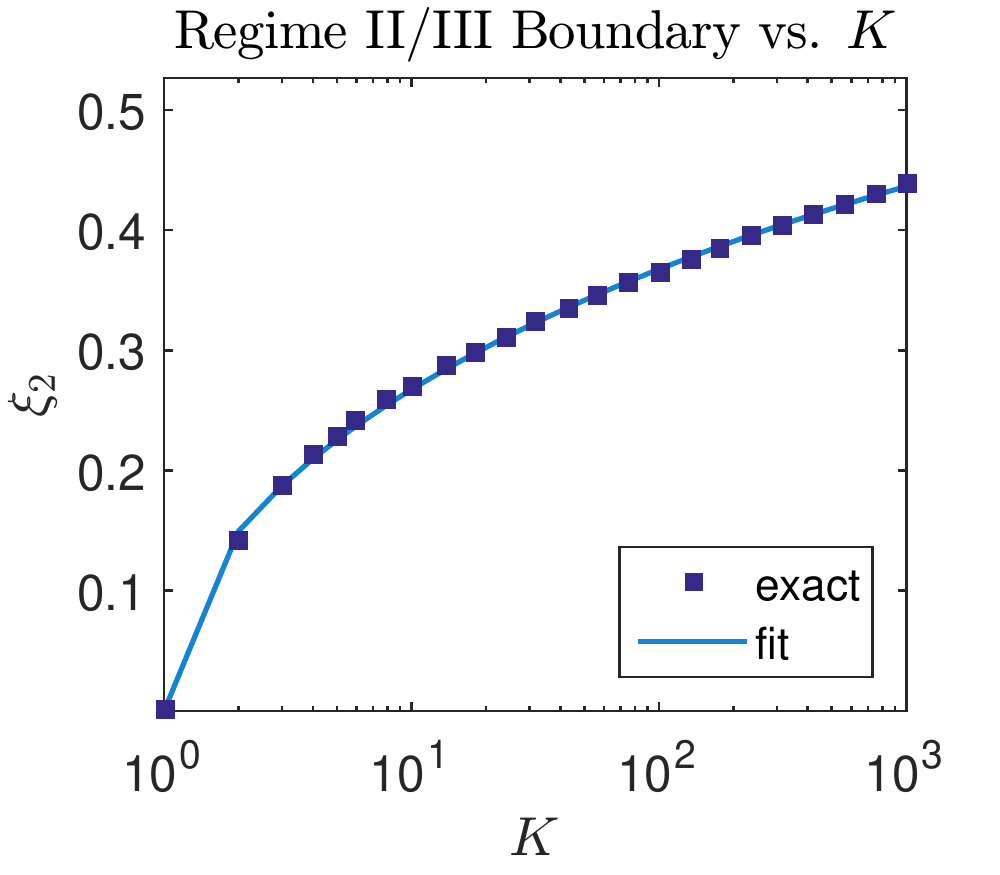}
        \caption{}
        \label{fig:xi2}
    \end{subfigure}
    \caption{The value of $\xi_1$ substantially decreases and $\xi_2$ slowly increases as $K$ increases, expanding Regime~II\@. (a) A log-log-cubic fit can be used to compute a close approximation to $\xi_1$ for all $K$, while a log-log-linear fit suffices for $K>20$. (b) The square root of a log-quadratic fit closely approximates $\xi_2$.}
    \label{fig:xi12}
\end{figure}

We define $\xi_1$ as the value of $\sigD$ where NMSE(mid) and $\nvargg$ intersect, which from~\eqref{eq:nmse_mid} and~\eqref{eq:asympGGML} is the solution to
\begin{equation}\label{eq:xi_1}
    \beta(\phat)[(\sigD)^2+1/12] = \frac{K/2 }{K^2+3K+2}
\end{equation}
for a given $K$. We remind the reader that $\phat$ is also dependent on $\sigD$ as shown in~\eqref{eq:match_kurt}.
Fig.~\ref{fig:xi1} shows that $\xi_1$ decreases as $K$ increases, since the probability of observing an ``outlier'' measurement due to the exponential tails increases with $K$, so a lower $\sigD$ value (i.e., with shorter tails) is needed for the midrange estimator to achieve nearly-optimal performance.
The figure shows the exact values of $\xi_1$ computed by solving~\eqref{eq:xi_1} as well as a log-log-cubic least-squares fit
\begin{align}\label{eq:xifit_llc}
    \log{\xi}_1 &\approx 0.0104(\log{K})^3-0.1760(\log{K})^2 \nonumber \\
    & \quad+0.0274(\log{K})-1.8511,
\end{align}
which can be used quickly to calculate an approximation for a desired value of $K$.
Since the relationship appears fairly linear for $K>20$, the simple log-log-linear fit
\begin{equation}\label{eq:xifit_ll}
    \log{\xi}_1 \approx -0.9301(\log{K})-0.1963,
\end{equation}
which can be rewritten as $\xi_1 \approx 0.8217/K^{0.9301}$,
is also useful for quick computation.
The natural logarithm is used in each case.

\subsection{Defining $\xi_2$}
Since NMSE(mean) and $\nvargg$ both have $1/K$ factors, they converge where $\beta(\phat) = 1$, which is only the case for $\phat=2$.
This suggests that equality requires the noise to be exactly Gaussian, which only occurs for $\sigD\rightarrow\infty$.
Instead, we can look for a point where
$\nvargg$ and NMSE(mean)
can reasonably be considered to have converged
(i.e., the GG is close enough to a Gaussian).

We propose that
a reasonable definition of $\xi_2$ is the value of $\sigD$ that minimizes NMSE(Q), the expected normalized MSE of $\muhatQ$.
Intuitively, as $\sigD$ increases from $\xi_2$, the Gaussian variance will dominate for both quantized and dithered measurements, so that the effect of the quantization error is negligible, whether signal-independent for dithered measurements or signal-dependent without dither.
Thus the point at which NMSE(Q) is minimized indicates where the Gaussian variance begins to dominate and is a reasonable place to consider a GG approximation to be sufficiently Gaussian.
We derive in Appendix~\ref{sec:apndx_qmse} that NMSE(Q) is given as 
\begin{align}\label{eq:qmse1}
\text{NMSE}&\text{(Q)} = \E[(\muhatQ - \muX)^2]/\Delta^2  \nonumber\\
=&\frac{1}{12} + \frac{1}{K} \int_{-1/2}^{1/2}  \sum_{m=-M}^M m^2 \Psimu d\muX\nonumber\\
& + \frac{K-1}{K} \int_{-1/2}^{1/2}  \left(\sum_{m=-M}^M m \Psimu\right)^2 d\muX\nonumber\\
& - 2 \int_{-1/2}^{1/2} \muX \sum_{m=-M}^M m \Psimu d\muX,
\end{align}
where 
\begin{equation}
    \Psimu = \Phi\!\left( \frac{m+1/2-\muX}{\sigD} \right) - \Phi\!\left(\frac{m-1/2-\muX}{\sigD}\right).
\end{equation}
Defining 
\begin{equation}
    \xi_2 = \argmin_{\sigD} \E[(\muhatQ - \muX)^2]/\Delta^2
\end{equation}
and solving via a Nelder-Mead algorithm~\cite{Nelder1965} and numerical integration, we show in Fig.~\ref{fig:xi2} that the value of $\xi_2$ changes only slightly as a function of $K$.

This range of values is notably very close to the value $\sigD=1/2$ recommended by Vardeman and Lee~\cite{Vardeman2005}, or the value $\sigD=1/3$ at which Moschitta et al. suggest that the loss of information from quantizing samples of a Gaussian distribution becomes negligible in estimation of the mean~\cite{Moschitta2015}.
For quick computation, $\xi_2$ can be approximated by the square root of a log-quadratic fit:
\begin{equation}
    \xi_2 \approx \sqrt{-0.000756(\log K)^2+0.328\log K}.
\end{equation}

We notice that the Regime boundary definitions are inconsistent for $K<3$, as $\xi_1>\xi_2$; however, the Regimes are meaningless for $K =$ 1 or 2 anyway, as symmetric order statistics-based estimators (e.g., mean, median, midrange) are all equivalent for such small numbers of measurements, so there is no advantage to distinguishing between noise distributions.
We notice also that since $\xi_1$ decreases monotonically and $\xi_2$ increases monotonically with $K$, Regime~II grows as $K$ increases, since small mismatches between the assumed and true PDFs become easier to observe.
Intuitively, $\xi_1$ decreases much faster than $\xi_2$ increases because the difference between a PDF with finite support ($\sigD = 0$) and one with infinite support ($\sigD>0$) is more significant for large $K$ than the difference between finite $\sigD$ (e.g., GG approximation with $\phat>2$ and $\sigD\rightarrow\infty$ ($\phat=2$).

\section{Numerical Results}\label{sec:results}

\begin{figure*}[t]
    \centering
    \begin{subfigure}{0.28\linewidth}
        \centering
        \includegraphics[trim={0cm 0cm 5.6cm 0cm},clip, width=\linewidth]{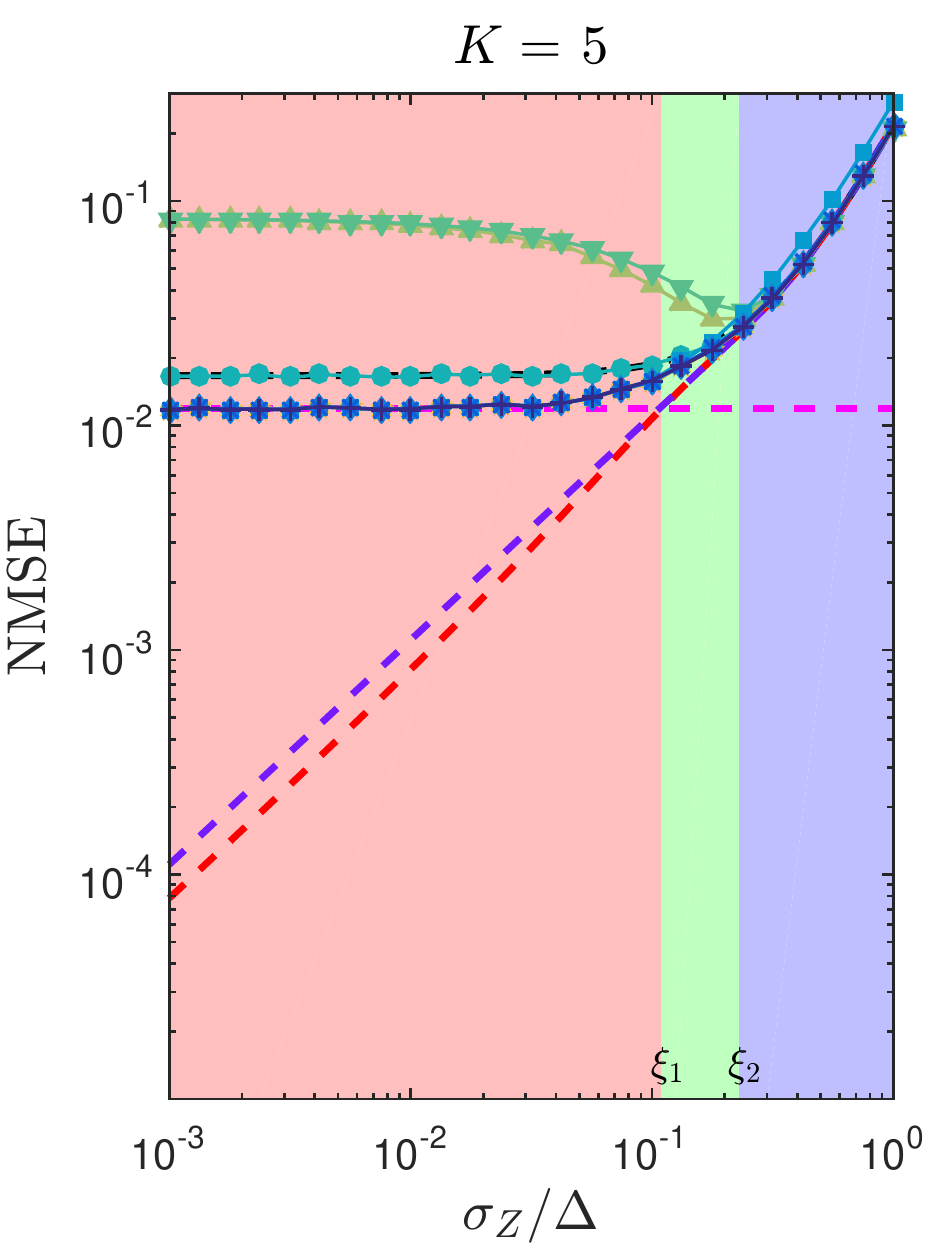}
        \caption{}
        \label{fig:sigZDelta_a}
    \end{subfigure}
    \hfill
    \begin{subfigure}{0.28\linewidth}
        \centering
        \includegraphics[trim={0cm 0cm 5.6cm 0cm},clip, width=\linewidth]{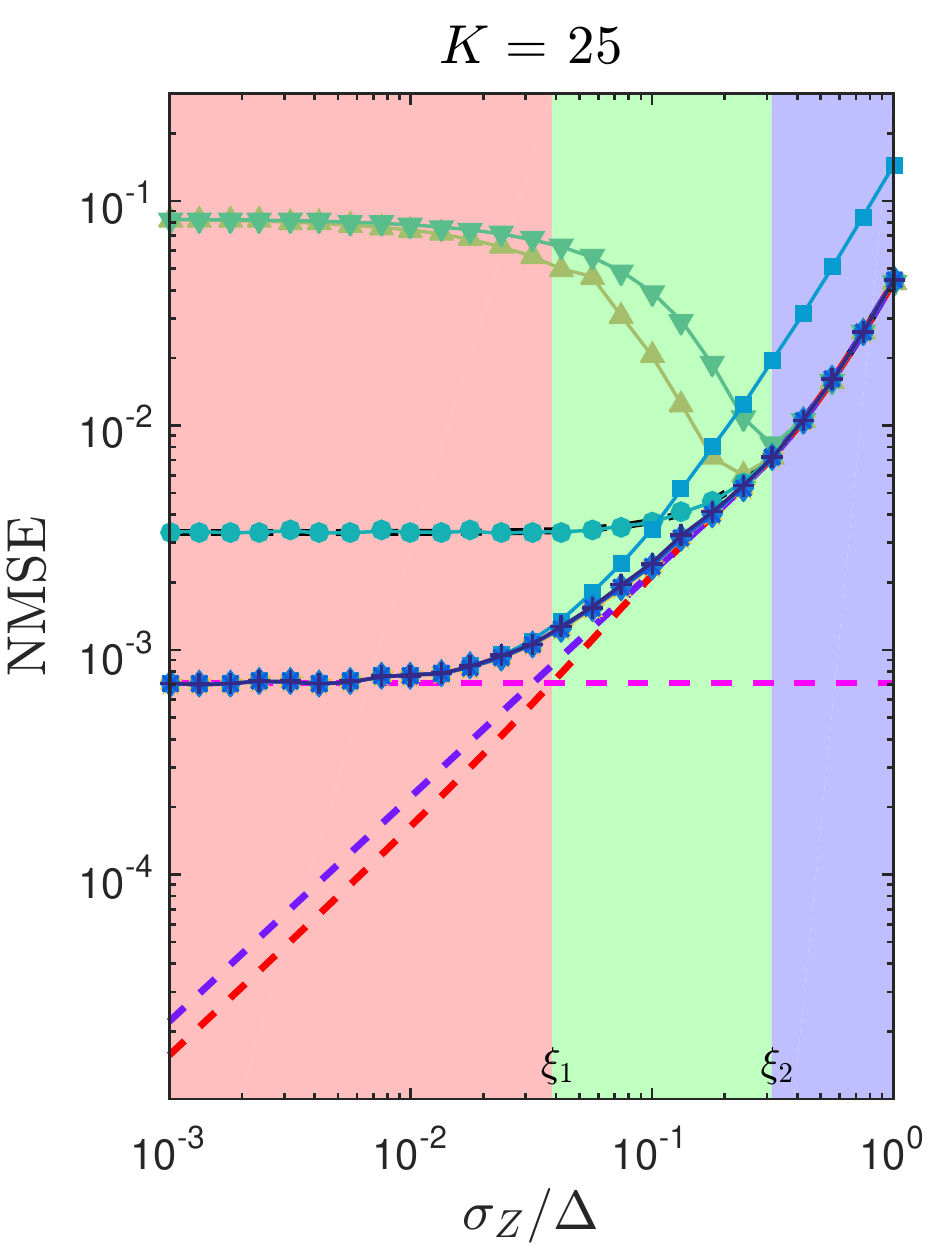}
        \caption{}
        \label{fig:sigZDelta_b}
    \end{subfigure}
    \hfill
    \begin{subfigure}{0.28\linewidth}
        \centering
        \includegraphics[trim={0cm 0cm 5.6cm 0cm},clip, width=\linewidth]{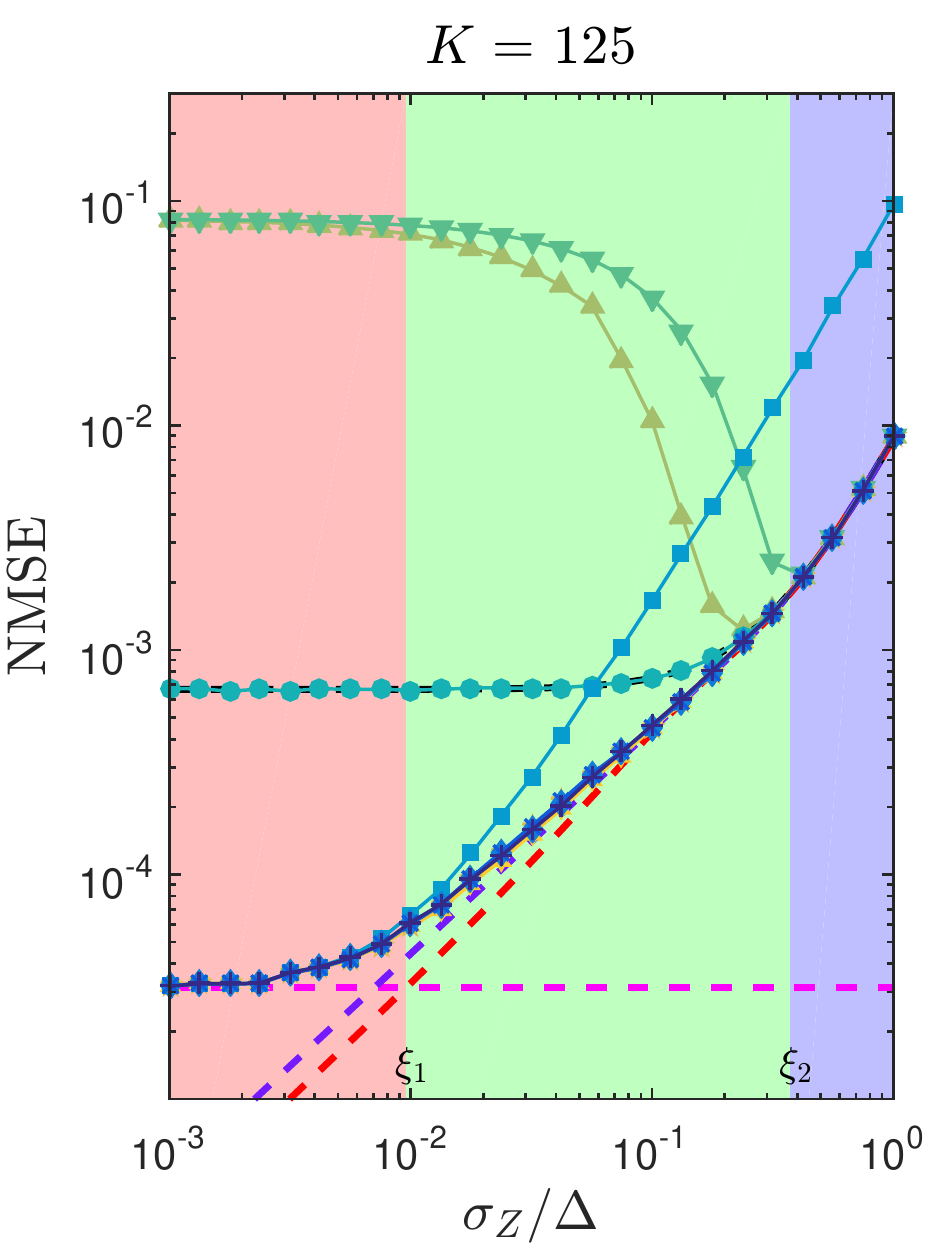}
        \caption{}
        \label{fig:sigZDelta_c}
    \end{subfigure}
    \hfill
    \begin{subfigure}{0.14\linewidth}
        \centering
        \includegraphics[trim={9.7cm 0cm 0cm 0cm},clip, width=\linewidth]{figures/rapp9}
    \end{subfigure}
    \caption{The performance of the estimators is evaluated for $K = $ (a) 5, (b) 25, and (c) 125 to show the range of behavior as $\sigD$ varies. The ML estimator for dithered measurements $\muhatDML$ and the estimators based on the GGD ($\muhatGGML$, $\muhatNB$, $\muhatNL$, and $\muhatAlpha$) achieve the lowest NMSE for each $\sigD$ regime (curves are overlapping). Results are shown for 20000 Monte Carlo trials.}
    \label{fig:perf_sigZ_vs_Delta}
\end{figure*}
Monte Carlo simulations were performed to compare the NMSE performance of the generalized Gaussian and order statistics-based estimators ($\muhatNB$, $\muhatNL$, $\muhatAlpha$) against the ML estimators ($\muhatDML$, $\muhatGGML$) and the conventional sample mean ($\muhatMean$) and midrange ($\muhatMid$).
Estimates were also computed applying the sample mean ($\muhatQ$) and ML estimator ($\muhatQML$) to the quantized data to determine under which conditions subtractive dithering actually provides an advantage. 
As in the motivating example in Section~\ref{sec:formulation}, for each Monte Carlo trial, $\muX$ was chosen uniformly at random from $[-\Delta/2,\Delta/2]$, and $K$ samples of signal noise $Z\sim\mathcal{N}(0,\sigmaZ^2)$ and dither $D\sim\uniform{-\Delta/2}{\Delta/2}$ were generated for~\eqref{eq:model}.
The quantization bin size was maintained at $\Delta = $ 1 throughout.
The normalized MSE was computed for $T = $ 20,000 trials. 

\begin{figure}
    \centering
    \includegraphics[trim={4mm 5mm 3mm 2mm},clip,width=\linewidth]{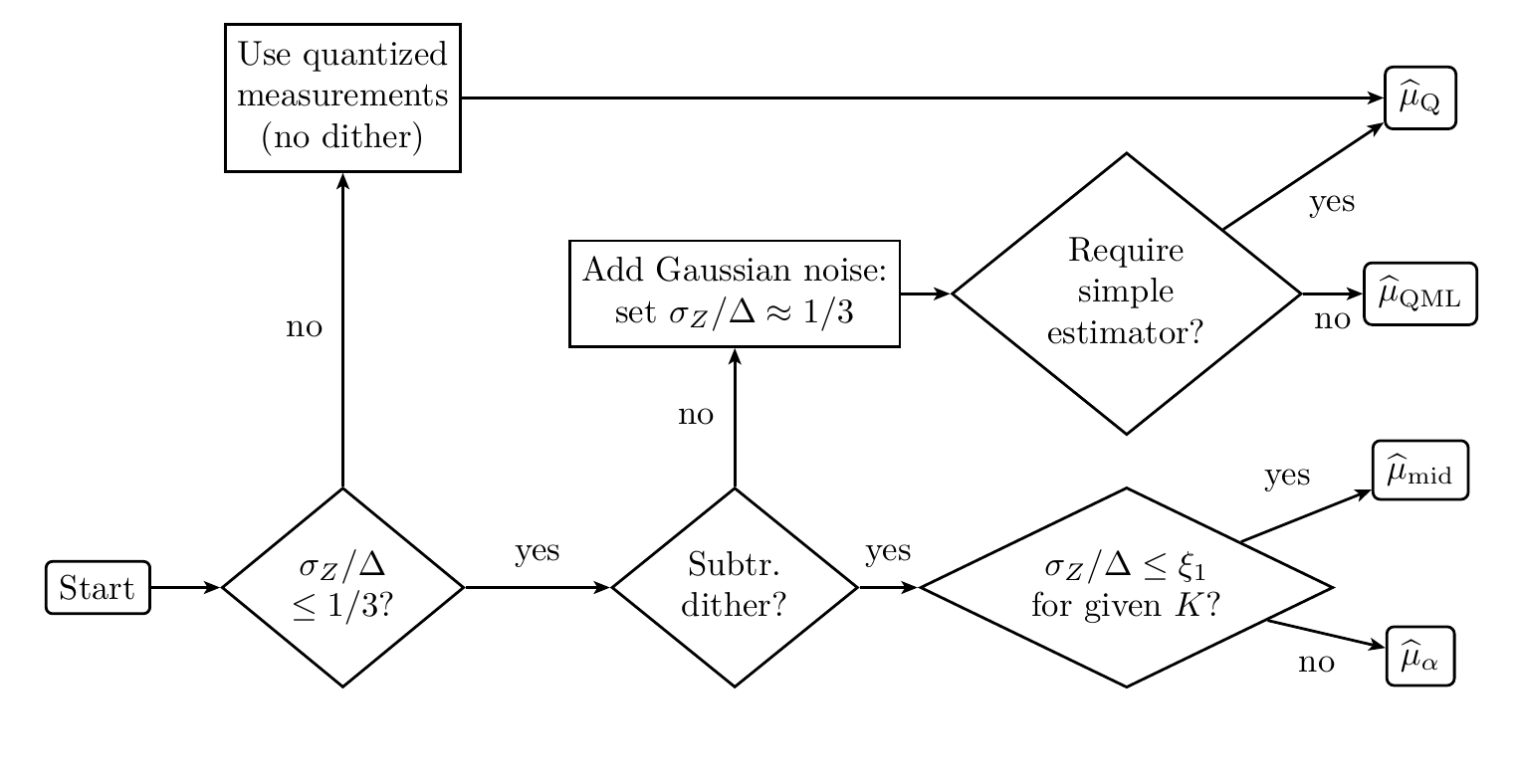}
    \caption{
    The results of our Monte Carlo simulations lead to a simplified decision process for when and how to use dither. 
    If $\sigD>\xi_2 ~(\approx1/3)$, there is no benefit to using anything but $\muhatQ$ applied to the quantized measurements, whereas dither leads to reduce estimation error when $\sigD \leq 1/3$.
    If subtractive dithering is not possible, the best performance can be achieved by adding Gaussian noise to set $\sigD \approx 1/3$ and applying $\muhatQML$ (although $\muhatQ$ can be used if simplicity is required).
    However, larger performance improvements can be achieved with a subtractively-dithered quantizer.
    For $K$ subtractively-dithered measurements, compute $\xi_1$ from  either~\eqref{eq:xifit_llc} or~\eqref{eq:xifit_ll} to determine whether to use $\muhatMid$ (in Regime~I) or $\muhatAlpha$ (in Regime~II).}
    \label{fig:flowchart}
\end{figure}

\begin{figure*}[t]
    \captionsetup{belowskip=2pt}
    \centering
    \begin{subfigure}{0.32\linewidth}
        \centering
        \caption*{$\sigD=0.004$}
        \includegraphics[trim={0mm 4.4mm 2.5mm 10.5mm},clip, width=\textwidth]{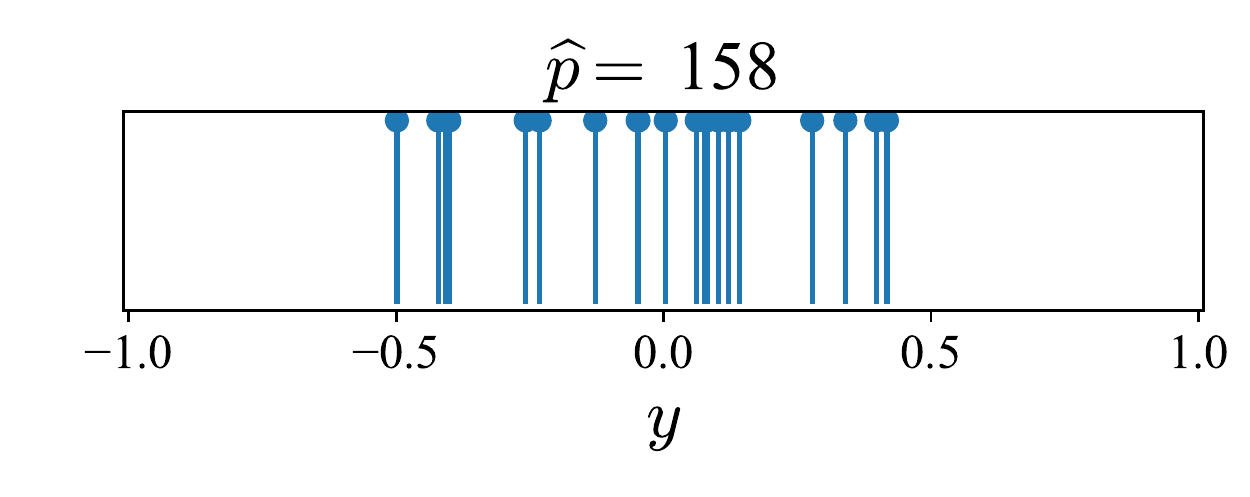}
        \caption{}
        \label{fig:osc_a}
    \end{subfigure}
    \hfill
    \begin{subfigure}{0.32\linewidth}
        \centering
        \caption*{$\sigD=0.04$}
        \includegraphics[trim={0mm 4.4mm 2.5mm 10.5mm},clip, width=\textwidth]{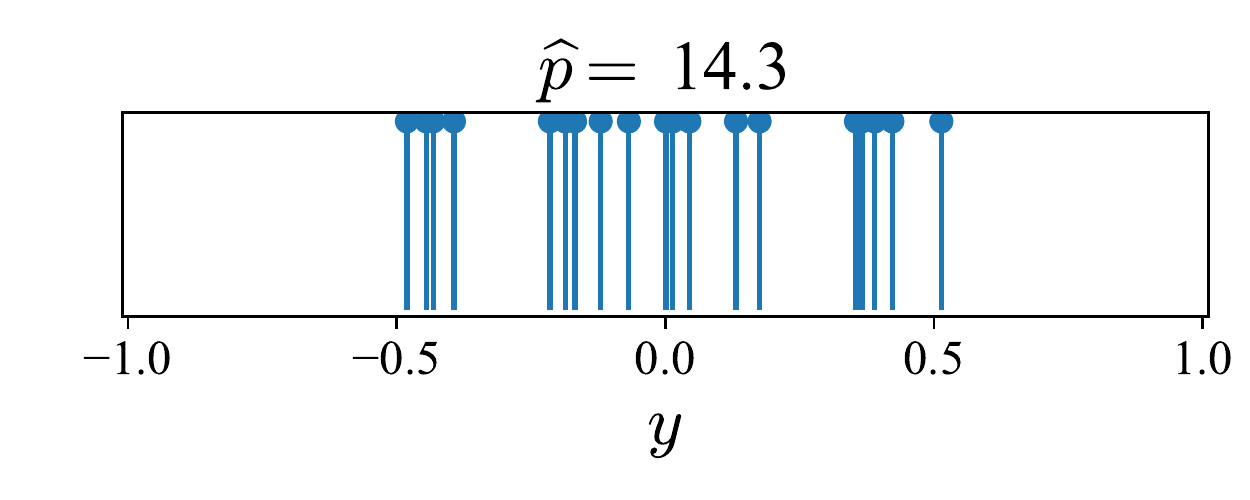}
        \caption{}
        \label{fig:osc_b}
    \end{subfigure}
    \hfill
    \begin{subfigure}{0.32\linewidth}
        \centering
        \caption*{$\sigD=0.4$}
        \includegraphics[trim={0mm 4.4mm 2.5mm 10.5mm},clip, width=\textwidth]{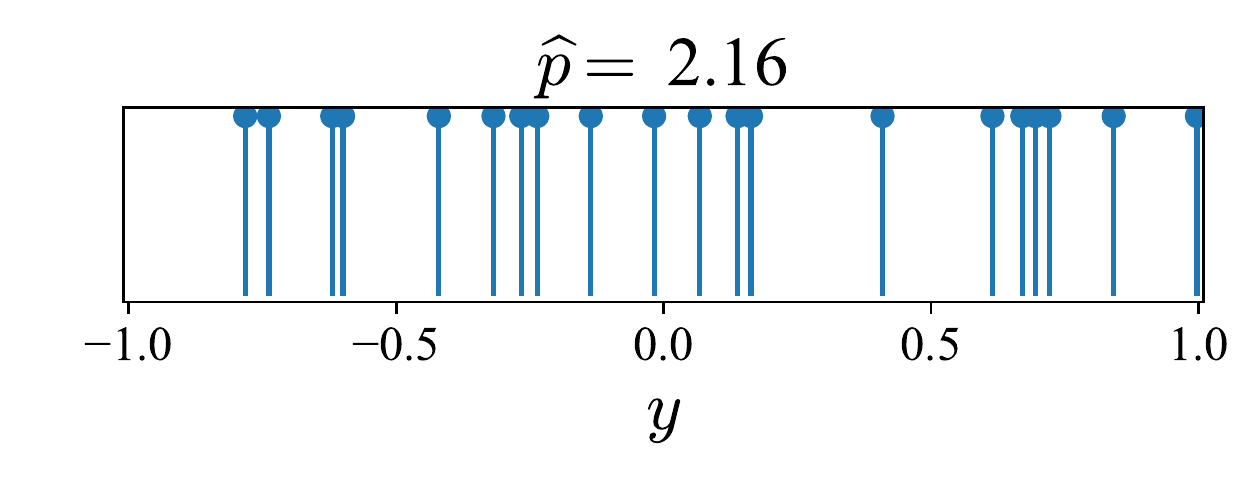}
        \caption{}
        \label{fig:osc_c}
    \end{subfigure}
    \\
    \begin{subfigure}{0.32\linewidth}
        \centering
        \caption*{$\phat=158$}
        \includegraphics[trim={0cm 4.8mm 0.25cm 0.3cm},clip, width=\textwidth]{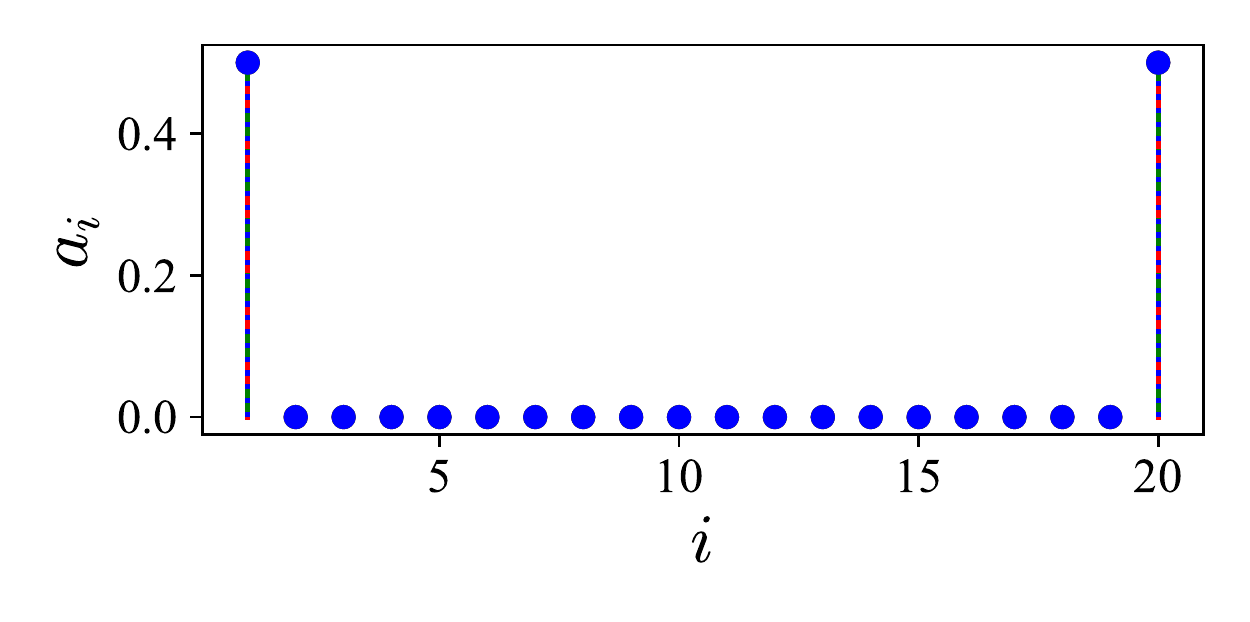}
        \caption{}
        \label{fig:osc_d}
    \end{subfigure}
    \hfill
    \begin{subfigure}{0.32\linewidth}
        \centering
        \caption*{$\phat=14.3$}
        \includegraphics[trim={0cm 4.8mm 0.25cm 0.3cm},clip, width=\textwidth]{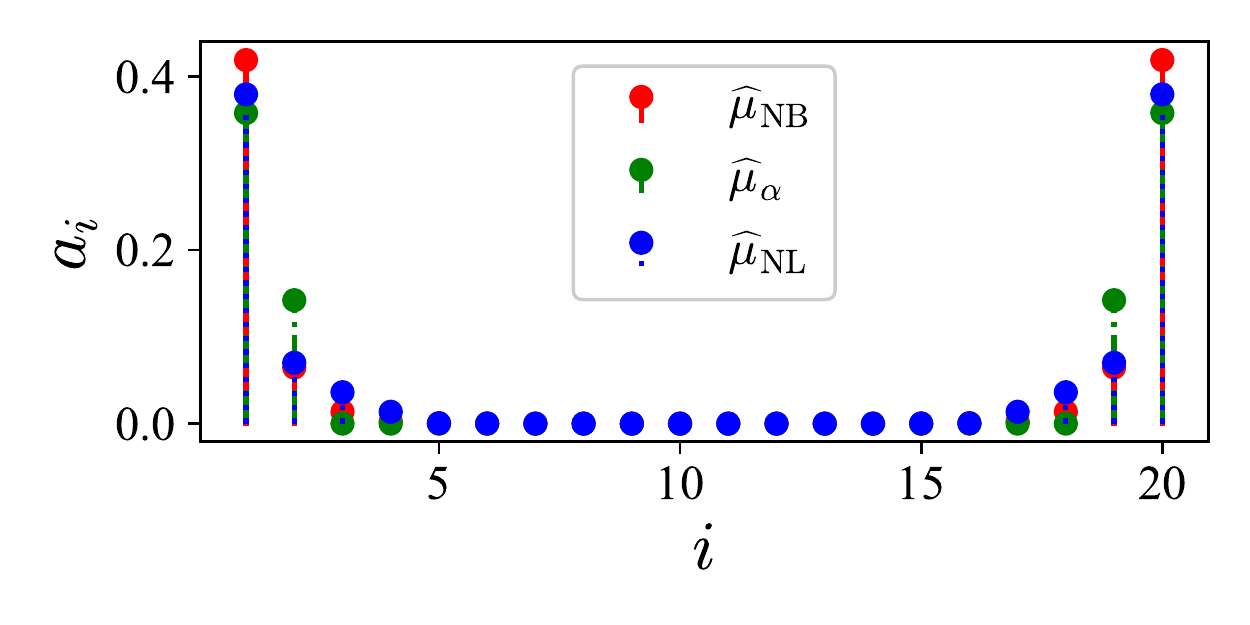}
        \caption{}
        \label{fig:osc_e}
    \end{subfigure}
    \hfill
    \begin{subfigure}{0.32\linewidth}
        \centering
        \caption*{$\phat=2.16$}
        \includegraphics[trim={0cm 4.8mm 0.25cm 0.3cm},clip, width=\textwidth]{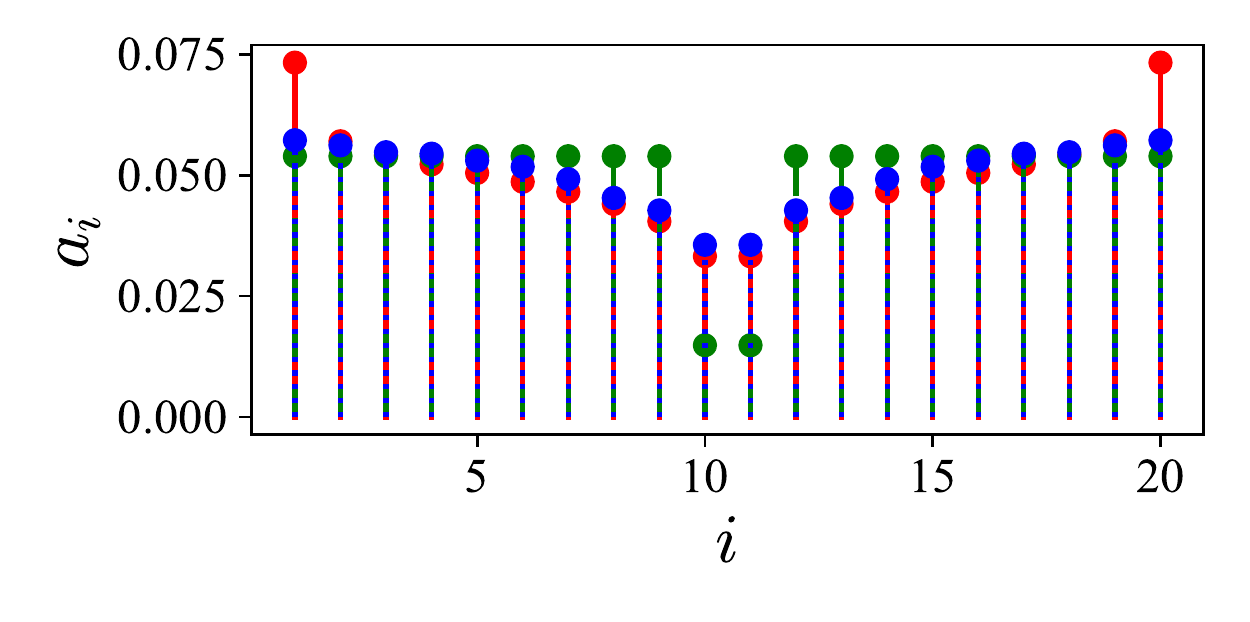}
        \caption{}
        \label{fig:osc_f}
    \end{subfigure}
    \\
    \begin{subfigure}{0.75\linewidth}
        \centering
        \caption*{$\phat=14.3,~K=100$}
        \includegraphics[trim={0.25cm 4.8mm 0.25cm 0.3cm},clip, width=\textwidth]{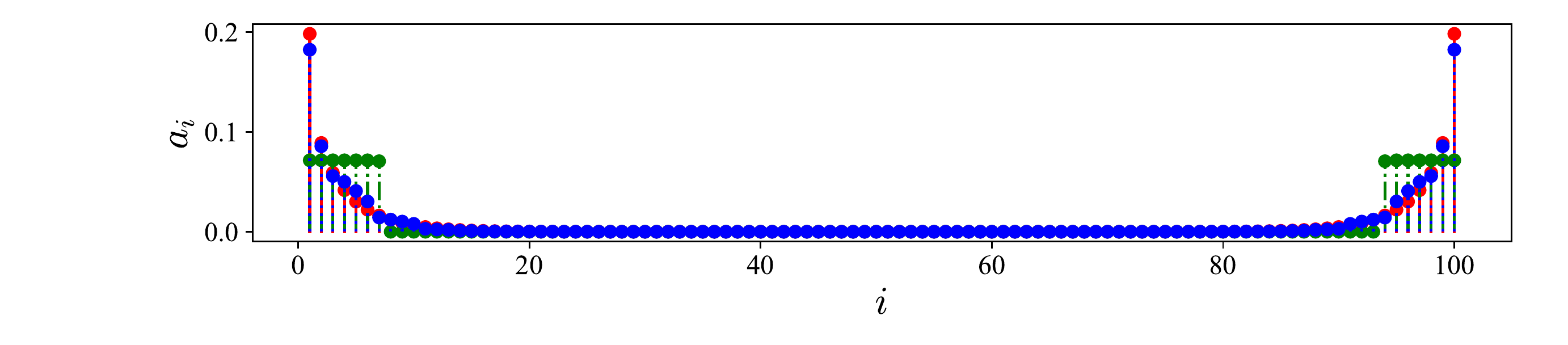}
        \caption{}
        \label{fig:osc_g}
    \end{subfigure}
    \caption{Example dithered measurements are shown in (a-c) for $\sigD = $ 0.004, 0.04, and 0.4 with $K=20$. Plots (d-f) show the resulting coefficient values for the GG estimators given the estimated value of $\phat$ above. In (g), $\sigD$ = 0.04 and $K$ = 100, highlighting how the coefficients change as $K$ increases. Note that the coefficients of the order statistics for the NL estimator depend on the measured data sequence shown above.}
    \label{fig:ord_stats_coeffs}
\end{figure*}

\begin{figure*}[t]
    \centering
    \begin{subfigure}{0.28\linewidth}
        \centering
        \includegraphics[trim={0cm 0cm 58mm 0cm},clip, width=\textwidth]{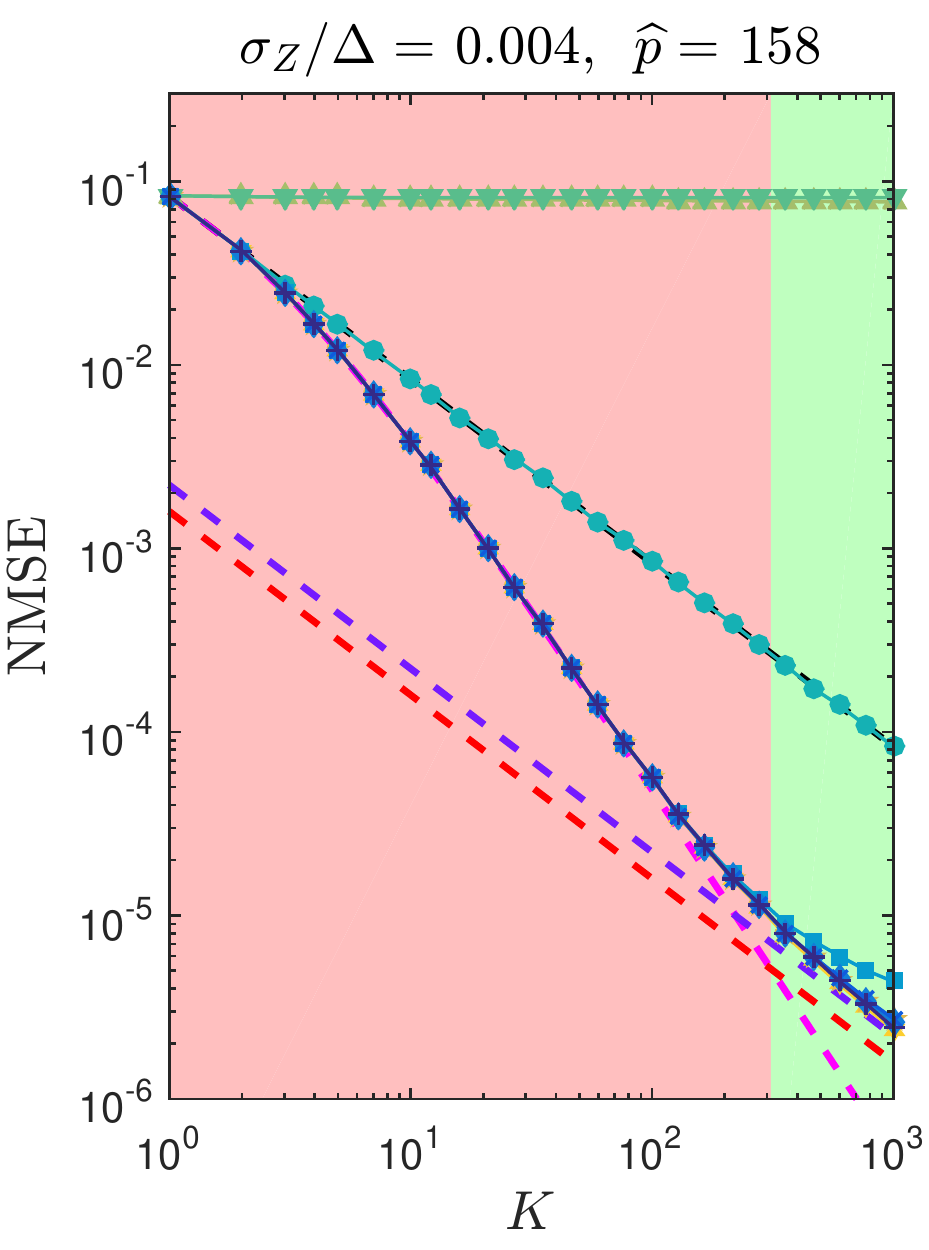}
        \caption{}
        \label{fig:res_a}
    \end{subfigure}
    \hfill
    \begin{subfigure}{0.28\linewidth}
        \centering
        \includegraphics[trim={0cm 0cm 58mm 0cm},clip, width=\textwidth]{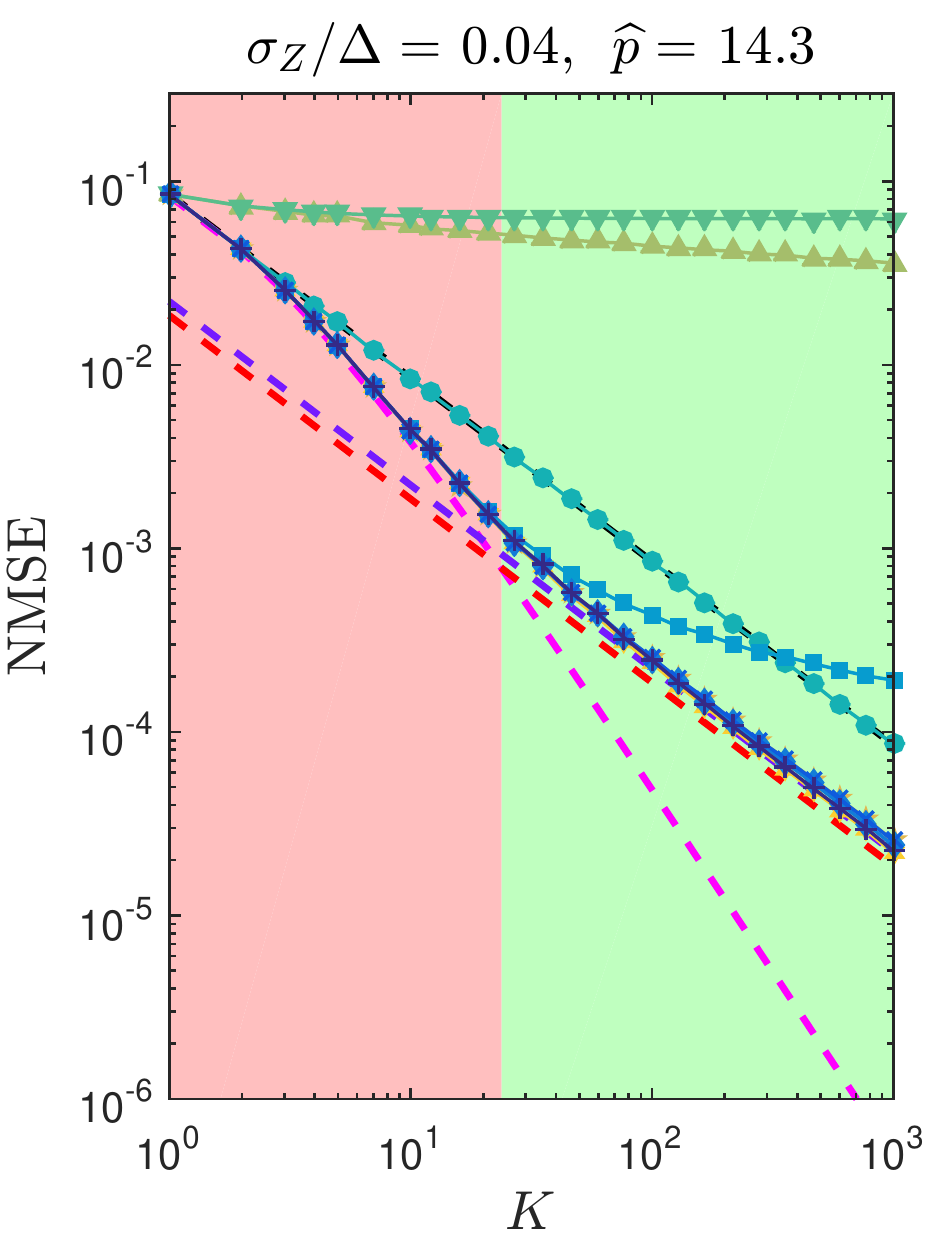}
        \caption{}
        \label{fig:res_b}
    \end{subfigure}
    \hfill
    \begin{subfigure}{0.28\linewidth}
        \centering
        \includegraphics[trim={0cm 0cm 58mm 0cm},clip, width=\textwidth]{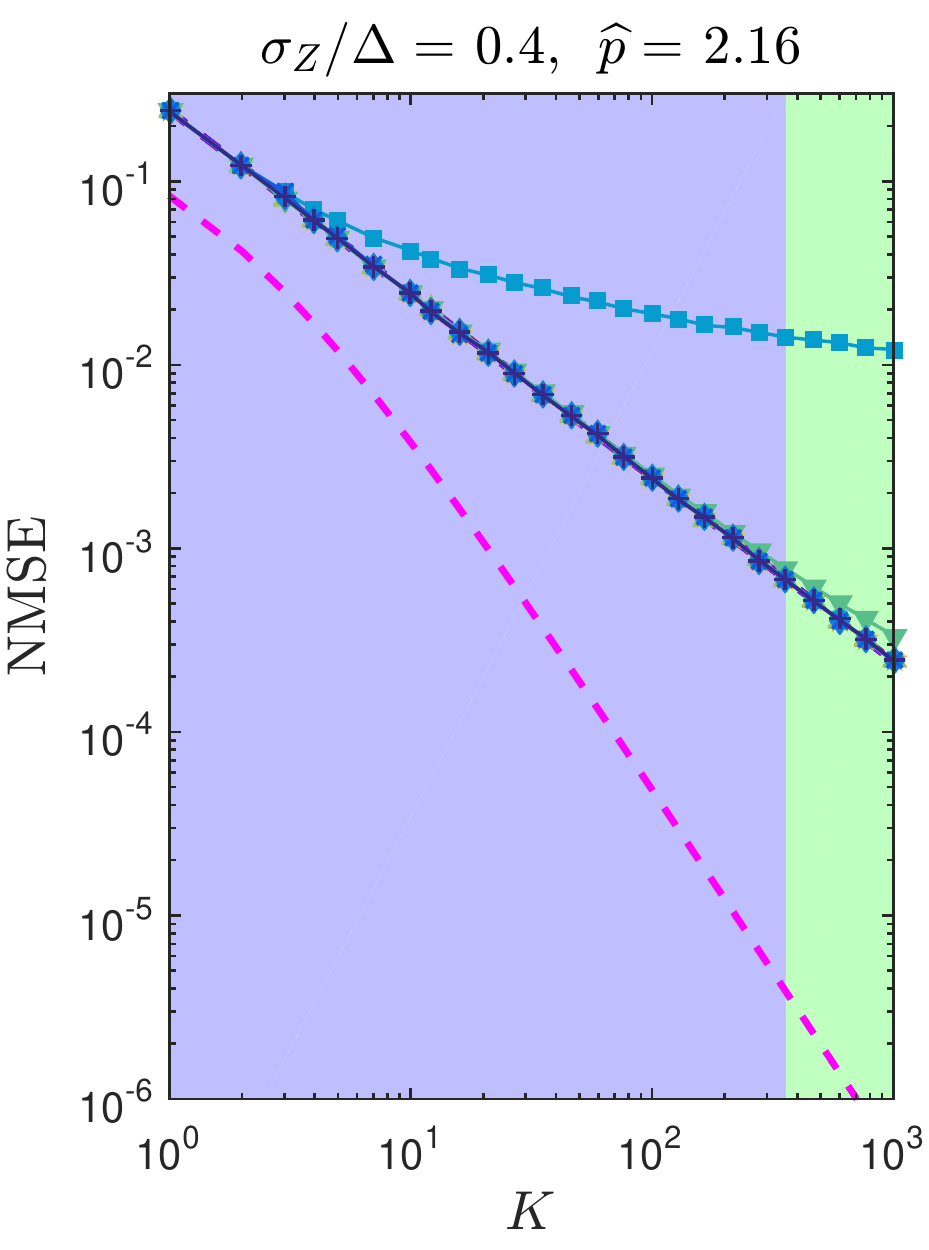}
        \caption{}
        \label{fig:res_c}
    \end{subfigure}
        \hfill
    \begin{subfigure}{0.14\linewidth}
        \centering
        \includegraphics[trim={9.7cm 0cm 0cm 0cm},clip, width=\linewidth]{figures/rapp9}
    \end{subfigure}
    \caption{The performance of the order statistics estimators is evaluated for $\sigD = $ (a) 0.004, (b) 0.04, and (c) 0.4 to show the full range of behavior as the number of measurements $K$ increases. The plots show the MSE normalized to $\Delta = 1$ from 20000 trials per data point. Dashed lines show the theoretical NMSE of the mean and midrange, and the asymptotic variance of the ML estimator of the GGD mean.
     The ML estimator for dithered measurements $\muhatDML$ and the estimators based on the GGD ($\muhatGGML$, $\muhatNB$, $\muhatNL$, and $\muhatAlpha$) achieve the lowest NMSE for all $K$ (curves are overlapping).
    }
    \label{fig:perf_Delta_vs_sigZ}
\end{figure*}

\subsection{Normalized MSE vs. $\sigD$}
We begin by discussing the plots in Fig.~\ref{fig:perf_sigZ_vs_Delta} of NMSE as a function of $\sigD$ for $K =$ 5, 25, and 125.
Because nine separate estimators and four NMSE bounds are displayed in each plot, we acknowledge that the figures can be difficult to follow due to the overlapping curves.
A flowchart is included in Fig.~\ref{fig:flowchart} that summarizes the results and provides a decision-making process for whether to use dither and which estimator to choose.

\subsubsection{Generalized Gaussian Estimators}
The GG-based estimators ($\muhatNB$, $\muhatNL$, $\muhatAlpha$, $\muhatGGML$) have effectively identical performance and match that of $\muhatDML$.
The actual differences in performance vary on the order of a few percent over large ranges of $K$ and $\sigD$, compared to the orders of magnitude differences for the mean and midrange.
The negligible performance difference further validates approximating the total noise with the GGD\@.
For this same reason, we collectively discuss $\muhatDML$ and the GG-based estimators in the following sections.

The GG-based estimators meet or exceed the performance of all other estimators for all $\sigD$ and for all $K$.
More specifically, the GG estimators converge to and match the performance of the midrange in Regime~I and likewise converge to and match the performance of the mean in Regime~III\@.
In Regime~II, the GG estimators outperform both the mean and the midrange.
Thus, a GG estimator should be the default estimator choice for any $\sigD$.

Given the approximate equivalence of the GG estimators, we argue that the trimmed-mean $\muhatAlpha$ is the best choice of general-purpose estimator for dithered data. 
The other estimators either require iterative solvers ($\muhatDML$, $\muhatGGML$), rely on numerical computation for the GG inverse CDF ($\muhatNB$), or are data-dependent ($\muhatNL$).
On the other hand, $\muhatAlpha$ has a simple closed-form solution that can be tabulated if needed.

\subsubsection{Performance by Regime---Dithered Measurements}
The plots in Fig.~\ref{fig:perf_sigZ_vs_Delta} validate the concept of three distinct regimes of noise behavior.
In the plots, the approximate regime boundaries are computed to be $\xi_1 = \{0.1098 , 3.85\times10^{-2}, 9.56\times10^{-3} \}$ and $\xi_2 = \{0.2296, 0.3132, 0.3737\}$  for $K = \{5, 25, 125\}$, respectively,
confirming that Regime~II expands as $K$ increases.
In Regime~I, the NMSE performance of all estimators on the dithered data is basically flat and equal to NMSE(mid). 
This suggests that for a practical system where $\sigD$ can be tuned, once the system is operating in Regime~I (dependent on a fixed $K$), there is no benefit from further decreasing $\sigD$; performance can only be improved by increasing $K$.
In Regime~II, the GG-based estimators approach NCRB($\muX$), especially for large $K$.
We note that while $\nvargg$ and NCRB($\muX$) are close in Regime~II, NCRB($\muX$) is a tighter bound, as it is based on the true noise distribution, although $\nvargg$ may be easier to compute for a rough estimate of performance.
In Regime~III, the NMSE performance of all estimators on the dithered data is equal to NMSE(mean).
In both Regimes II and III, the NMSE decreases as $\sigD$ decreases.
Performance likewise improves with increasing $K$.

\subsubsection{Performance by Regime---Quantized Measurements}
While the three Regimes were technically defined for dithered measurements in particular, they are also informative of the behavior of estimators applied to quantized measurements.
In Regime~I, $\sigD$ is so small that, unless $\muX$ lies on the boundary between quantization bins, all measurements are quantized to the same value.
As a result, the NMSE of both $\muhatQ$ and $\muhatQML$ is dominated by the squared bias term, which is $1/12$ (the variance is zero).
Further decreasing $\sigD$ or increasing $K$ provides no benefit.

In Regime~II, $\sigD$ is large enough that there is often some variation in the measurements due to signal even without the addition of dither.
This phenomenon is sometimes referred to in the literature as self-dithering, equivalent to adding nonsubtractive Gaussian dither to a constant signal $\muX$~\cite{Carbone1994}.
Within Regime~II, both $\muhatQ$ and $\muhatQML$ improve as $\sigD$ increases because the increased signal variation reduces the bias term of the NMSE faster than the variance increases.
The NMSE is minimized for $\muhatQ$ by definition at $\xi_2$, and then the NMSE increases as $\sigD$ increases in Regime~III\@.
This suggests that if $\sigD$ is small and subtractive dither cannot be used, then quantized measurements benefit from adding nonsubtractive Gaussian dither such that $\sigD=\xi_2$, which is approximately $1/3$.
It is in Regime~II that $\muhatQML$ shows the largest improvement in performance over $\muhatQ$,
with the ML estimator accounting for the form of the signal variation for quantized measurements.

In Regime~III, the NMSE of $\muhatQ$ and $\muhatQML$ matches that of the best estimators applied to dithered data.
Clearly, $\sigD$ is large enough that even the quantized measurements contain sufficient information about the signal variation.
This suggests that dither provides no benefit in Regime~III, since equal performance can be achieved without dither.
Again for both Regimes II and III, the NMSE decreases as $K$ is increased.

\subsection{Order Statistics-Based Estimator Coefficients}
To better understand why the order statistics-based estimators have essentially identical performance,
in Fig.~\ref{fig:ord_stats_coeffs}
we plot the coefficients $a_i$ from~\eqref{eq:l-est} for each estimator.
The top row shows example measurements for $K =$ 20, $\Delta=1$, and $\sigD = $ 0.004, 0.04, and 0.4, respectively, with the samples spreading out as the Gaussian variance increases.
The second row of plots depicts the resulting coefficients for $\muhatNB$, $\muhatNL$, and $\muhatAlpha$ using the estimated value $\phat$.
Fig.~\ref{fig:osc_d} shows the coefficients are equivalent to those of  $\muhatMid$ for small $\sigD$.
In Figs.~\ref{fig:osc_e} and~\ref{fig:osc_f}, the coefficients of the various estimators are no longer identical.
However, the coefficients follow the same trends for each estimator, with zero weight on the middle order statistics for small $\sigD$ and more evenly-distributed weights as the noise model approaches a Gaussian.
We note that the coefficients for $\muhatNL$ vary depending on the particular set of measurements shown in the top row, and that different sample realizations can result in coefficients more or less similar to those of $\muhatNB$ and $\muhatAlpha$.
To show the behavior of the coefficients as $K$ increases, we also plot $\{a_i\}_{i=1}^K$ for $K = $ 100 and $\sigD = $ 0.04 in Fig.~\ref{fig:osc_g}.
This plot underscores that the coefficients for $\muhatAlpha$ are basically indicators of the most significant non-zero coefficients of $\muhatNB$ and $\muhatNL$. 
Using the simple formulation of $\muhatAlpha$ as a guide, in the limit as $K\rightarrow \infty$, only $K\alpha$ coefficients would have nonzero weight.
Since the performance of all three order statistics-based estimators is similar, this further suggests that the selection of which order statistics are used is more important than exactly how much they are weighted.

\subsection{Normalized MSE vs. $K$}
To better understand how the number of measurements affects the estimators' performance, we plot results for three fixed values of $\sigD$ in each regime (0.004, 0.04, 0.4) while varying $K$ in Fig.~\ref{fig:perf_Delta_vs_sigZ}.

In Fig.~\ref{fig:res_a}, $\muhatMid$ follows NMSE(mid) as expected for Regime~I until $K\approx200$.
At that point, the NMSE of the midrange begins to diverge, with slower improvement as $K$ increases. 
Similarly, the GG estimators follow NMSE(mid) until $K\approx200$ and then switch to NCRB($\muX$).
This suggests that $\sigD = 0.004$ is in Regime~I for $K<200$ and in Regime~II for $K>200$.
This switch between regimes occurs near the intersection of NMSE(mid) and $\nvargg$ as a function of $K$, further validating these bounds as useful demarcations of estimator performance.
For all $K$ in the plotted range, the midrange and GG estimators outperform the mean.
The quantized estimators show almost no improvement as $K$ increases.

In Fig.~\ref{fig:res_b}, the midrange performance is similar to that in Fig.~\ref{fig:res_a}, with $\muhatMid$ following NMSE(mid) until the intersection of NMSE(mid) and $\nvargg$ and then improving more slowly as a function of the number of measurements, eventually being outperformed by $\muhatMean$ for large $K$.
The GG estimators likewise follow NMSE(mid) for small $K$ and switch to following NCRB($\muX$) after the intersection.
For large $K$, the NMSE of the GG estimators is a constant factor lower than that of $\muhatMean$, with this factor approximately given by $\beta(\phat)$.
The NMSE of the quantized estimators decreases slowly as $K$ increases, with marginally better performance for $\muhatQML$ than $\muhatQ$.

Figures~\ref{fig:res_a} and~\ref{fig:res_b} help answer the question of how the order statistics-based estimators ``between'' the midrange and the mean would perform as a function of $K$.
The results suggest that these estimators ultimately have $O(K^{-1})$ NMSE reduction, although this reduction is faster for small values of $K$.

In Fig.~\ref{fig:res_c}, the noise can be sufficiently described as Gaussian for $K<359$; however, for larger $K$, $\xi_2>0.4$ as shown in Fig.~\ref{fig:xi2}.
The midrange has poor performance for all $K$, while the other dithered estimators and the quantized estimators have essentially identical performance for $K<359$.
Those estimators follow NMSE(mean), $\nvargg$, and NCRB($\muX$), which have converged.
In this regime, it is clear that there is no benefit to using dither, as there is minimal improvement in performance even for large $K$.
In fact, implementing a dithered quantizer is likely more complicated in practice and is discouraged for Regime~III\@.
\section{Conclusion}\label{sec:conclusion}

This work studied the task of estimating the mean of a Gaussian signal from quantized measurements.
By applying subtractive dither to the measurement process, the noise becomes signal-independent but no longer has a Gaussian distribution.
We showed that the generalized Gaussian distribution is a close and useful approximation for the Gaussian plus uniform total noise distribution.
Estimators using the generalized Gaussian approximation effectively match the performance of the ML estimator for the total noise, which is a significant improvement over the conventional mean and midrange estimators in Regime~II\@. 
Due to its computational simplicity and efficient performance, we recommend the trimmed mean $\muhatAlpha$.
From further comparison against estimators for quantized measurements, we determined simple design rules for deciding whether and how to use quantized measurements.
In short, there is value in using dither in Regimes I and II, and a GG-based estimator should be used in Regime II\@.
Future work will address variations on the measurement model, including non-Gaussian signal distributions and different dither implementations.

\appendices
\section{Cram{\'e}r-Rao Bound}\label{sec:crb}

\def\fy{f_Y(y;\muX,\sigmaZ,\Delta)}
\def\ppm{\frac{\partial}{\partial \muX}}

The Cram{\'e}r-Rao Bound is a lower bound on the variance of an unbiased estimator~\cite{VanTrees2013}, given by
\begin{equation*}
    \text{CRB}(\muX) = 1/I(\muX),
\end{equation*}
where $I(\muX)$ is the Fisher information computed as
\begin{align}
    I(\muX)&=\E\left[\left(\frac{\partial \log \fy}{\partial \muX}\right)^2 \right] \nonumber \\
    &\overset{(a)}{=} \int \frac{\left(\ppm f_Y(y;\muX,\sigmaZ,\Delta)\right)^2}{f_Y(y;\muX,\sigmaZ,\Delta)} dy \nonumber \\
    &\overset{(b)}{=}\frac{1}{\sigmaZ^2\Delta} \int \frac{\left[\phi\left(\frac{v-\Delta/2}{\sigmaZ}\right)-\phi\left(\frac{v+\Delta/2}{\sigmaZ}\right) \right]^2}{\Phi\left(\frac{v+\Delta/2}{\sigmaZ}\right)-\Phi\left(\frac{v-\Delta/2}{\sigmaZ}\right)} dv \nonumber \\
    &\overset{(c)}{=}\frac{1}{\sigmaZ^2} \int \frac{\left[\phi\left(\frac{u-1/2}{\sigD}\right)-\phi\left(\frac{u+1/2}{\sigD}\right) \right]^2}{\Phi\left(\frac{u+1/2}{\sigD}\right)-\Phi\left(\frac{u-1/2}{\sigD}\right)} du,
\end{align}
where step ($a$) uses the definition of expectation and the chain rule, $(b)$ differentiates~\eqref{eq:true_pdf} with respect to $\muX$ for $v=y-\muX$, and ($c$) changes variables to $u = v/\Delta$.
Normalizing by $\Delta^2$ removes the separate dependence on $\sigmaZ$ or $\Delta$, so we define the normalized CRB as 
\begin{align}
    \text{NCRB}(\muX) &= \text{CRB}(\muX)/\Delta^2 \nonumber \\
    &= \frac{(\sigD)^2}{\ds \int  \frac{\left[\phi\left(\frac{u-1/2}{\sigD}\right)-\phi\left(\frac{u+1/2}{\sigD}\right) \right]^2}{\Phi\left(\frac{u+1/2}{\sigD}\right)-\Phi\left(\frac{u-1/2}{\sigD}\right)} du}.
\end{align}
Finally, Fisher information is additive for independent observations, so for $K$ independent samples, the lower bound on the NCRB is $\Frac{1}{K}$ times that for one observation.
\section{Kurtosis Matching}\label{sec:kurtosis}
\def\pv{p_v}
\def\px{p_x}
\def\pz{p_z}

\def\gamv{\gamma(V)}
\def\kapv{\kappa(V)}
\def\muv{\mu_V}
\def\mufourv{\mu_4(V)}
\def\sigv{\sigma_V}
\def\varv{\sigma_V^2}
\def\gamw{\gamma(W)}
\def\kapw{\kappa(W)}
\def\muw{\mu_W}
\def\mufourw{\mu_4(W)}
\def\sigw{\sigma_W}
\def\varw{\sigma_W^2}
\def\gamb{\gamma(B)}
\def\kapb{\kappa(B)}
\def\mub{\mu_B}
\def\mufourb{\mu_4(B)}
\def\sigb{\sigma_B}
\def\varb{\sigma_B^2}
\def\gamc{\gamma(C)}
\def\kapc{\kappa(C)}
\def\muc{\mu_C}
\def\mufourc{\mu_4(C)}
\def\sigc{\sigma_C}
\def\varc{\sigma_C^2}
\def\gama{\gamma(A)}
\def\kapa{\kappa(A)}
\def\mua{\mu_A}
\def\mufoura{\mu_4(A)}
\def\siga{\sigma_A}
\def\vara{\sigma_A^2}
\def\gamx{\gamma(X)}
\def\kapx{\kappa(X)}
\def\mux{\mu_X}
\def\mufourx{\mu_4(X)}
\def\sigx{\sigma_X}
\def\varx{\sigma_X^2}
\def\gamy{\gamma(Y)}
\def\kapy{\kappa(Y)}
\def\muy{\mu_Y}
\def\mufoury{\mu_4(Y)}
\def\sigy{\sigma_Y}
\def\vary{\sigma_Y^2}
\def\gamz{\gamma(Z)}
\def\kapz{\kappa(Z)}
\def\muz{\mu_Z}
\def\mufourz{\mu_4(Z)}
\def\sigz{\sigma_Z}
\def\varz{\sigma_Z^2}

The \emph{kurtosis} of a random variable $B$ is the standardized fourth central moment~\cite{Dudewicz1988}, defined as
\begin{equation}\label{eq:kurtosis_def}
    \kapb = \frac{\E[(B-\mub)^4]}{\{\E[(B-\mub)^2]\}^2} = \frac{\mufourb}{\sigb^4}.
\end{equation}
The \emph{excess kurtosis} $\gamb = \kapb-3$ is often used to simplify computations.
Define $A = B+C$, where $B$ and $C$ are independent random variables.
The kurtosis of the sum can be computed by expanding~\eqref{eq:kurtosis_def} as follows:
\begin{align}
    \kapa &= \frac{\E[(A-\mua)^4]}{\{\E[(A-\mua)^2]\}^2} = \frac{\E\{[(B-\mub)+(C-\muc)]^4\}}{\{\E[((B-\mub)+(C-\muc))^2]\}^2} \nonumber \\
    &= \frac{\mufourb+\mufourc+6\varb\varc}{(\varb+\varc)^2}, \nonumber
\end{align}
where independence eliminates the odd cross terms.
Then the excess kurtosis is 
\begin{align}\label{eq:exkurt}
    \gama &= \frac{\sigb^4\gamb+\sigc^4\gamc}{\siga^4}.
\end{align}
The kurtosis of Gaussian and uniform random variables is well-known and straightforward to compute from the definition; the excess kurtosis is 0 for a Gaussian and $-6/5$ for a uniform distribution.
From~\cite{Soury2015}, we have that the excess kurtosis%
\footnote{Note that the definition of kurtosis in~\cite{Soury2015} corresponds to the excess kurtosis in this work.}
of a GGRV $V$ with shape parameter $\pv$ is
\begin{equation}
    \gamv = \frac{\Gamma(1/\pv)\Gamma(5/\pv)}{[\Gamma(3/\pv)]^2} - 3.
\end{equation}
To fit the GGD to the sum of uniform and Gaussian random variables, we set the kurtosis of the approximation to match the kurtosis of the sum using~\eqref{eq:exkurt}
\begin{align}
    \frac{\Gamma(1/\pv)\Gamma(5/\pv)}{[\Gamma(3/\pv)]^2} &= 3+ \frac{\sigz^4\cdot 0 +\sigw^4(-6/5)}{(\varw+\varz)^2} \nonumber \\
    &= 3 - \frac{6}{5} \frac{1}{\left[1+12 \left(\frac{\sigz}{\Delta}\right )^2\right]^2},
\end{align}
where $\sigw^2 = \Frac{\Delta^2}{12}$.

\section{Mean Squared Error of $\muhatQ$}\label{sec:apndx_qmse}

We use iterated expectation to compute the MSE of $\muhatQ$ as
\begin{equation}
\E[(\muhatQ - \muX)^2] = \E\left[\E[(\muhatQ - \muX)^2\vert \muX]\right],
\end{equation}
with no prior knowledge on the true value so that we assume $\muX\sim\Um[-\Delta/2,\Delta/2]$ within a bin. 
Define a function $g:\mathbb{R}\to\mathbb{R}$ as $g(x):=\E[(\muhatQ - \muX)^2\vert \muX=x] $, then
\begin{align}
g(x)&= \E\left[\left(  \frac{1}{K} \sum_{i=1}^K q(x+Z_i) - x \right)^2\right]\nonumber\\
&= x^2 + \frac{1}{K^2}\left(\sum_{i=i}^K \E\left[ \left( q(x+Z_i) \right)^2\right]\right. \nonumber\\
&\hspace{5em} \left. + \sum_{i=1}^K\sum_{j\neq i} \E\left[q(x+Z_i)\right]\E\left[q(x+Z_j)\right]\right)\nonumber\\
&\qquad - \frac{2x}{K}\sum_{i=1}^K \E\left[q(x+Z_i)\right]\nonumber\\
&=x^2 + \frac{1}{K} \E\left[ \left( q(x+Z) \right)^2\right] +  \frac{ K-1}{K} \left(\E\left[q(x+Z)\right]\right)^2 \nonumber\\
&\qquad- 2x \E\left[q(x+Z)\right].\label{eq:gx}
\end{align}
Using the definition
\begin{equation}
    \Psim = \Phi\!\left( \frac{m+1/2-x}{\sigD} \right) - \Phi\!\left(\frac{m-1/2-x}{\sigD}\right),
\end{equation}
note that
\begin{align*}
\E&\left[q(x+Z)\right] \nonumber \\
&= \lim_{M\rightarrow\infty} \sum_{m=-M}^M m\Delta \int_{m\Delta-\Delta/2}^{m\Delta+\Delta/2} \frac{1}{\sigmaZ}\phi\left(\frac{z-x}{\sigmaZ}\right) dz\\
&\approx  \Delta \sum_{m=-M}^M m \Psim
\end{align*}
for some large number $M$.
Similarly,
\begin{equation*}
    \E\left[ \left( q(x+Z) \right)^2\right] \approx  \Delta^2 \sum_{m=-M}^M m^2 \Psim.
\end{equation*}
The MSE normalized by $\Delta^2$ then follows as
\begin{align}\label{eq:qmse}
\E[(\muhatQ &- \muX)^2]/\Delta^2 =  \nonumber\\
&\frac{1}{12} + \frac{1}{K} \int_{-1/2}^{1/2}  \sum_{m=-M}^M m^2 \Psim dx\nonumber\\
& + \frac{K-1}{K} \int_{-1/2}^{1/2}  \left(\sum_{m=-M}^M m \Psim\right)^2 dx\nonumber\\
& - 2 \int_{-1/2}^{1/2} x \sum_{m=-M}^M m \Psim dx.
\end{align}

\section*{Acknowledgment}
The authors would like to thank Dr. Yanting Ma for assistance with the derivations in Appendix~\ref{sec:apndx_qmse} and Dr. Dongeek Shin for helpful discussions on the role of nonsubtractive dither in improving estimation accuracy.
Computing resources provided by Boston University's Research Computing Services are also gratefully appreciated.

\bibliographystyle{IEEEtran}
\bibliography{ref,bibl}

\end{document}